\documentclass[%
 reprint,
 amsmath,amssymb,
 aps,
floatfix,
]{revtex4-2}

\usepackage[utf8]{inputenc} 
\usepackage{graphicx}
\usepackage{dcolumn}
\usepackage{bm}
\usepackage{amsmath}
\usepackage{enumitem}
\usepackage{hyperref} 
\usepackage[user]{zref}
\usepackage{float}
\usepackage{placeins}
\usepackage{multirow}
\usepackage{array}
\usepackage{makecell}
\usepackage{tabularray}
\usepackage{float}

\hypersetup{
    colorlinks=true,
    linkcolor=blue,
    citecolor=blue,
    filecolor=blue,
    urlcolor=blue,
}
\newcommand{\ve}[1]{\mathbf{#1}}
\newcommand{\te}[1]{\overline{\overline{\mathrm{#1}}}}

\newcommand{\figref}[2]{\hyperref[#1]{~Fig.~\ref*{#1}{#2}}}
\newcommand{\secref}[1]{Section~\ref{#1}}
\newcommand{\subsecref}[1]{subsection \ref{#1}}
\newcommand{\subsecrange}[3]{\ref{#1}--\ref{#2}--\ref{#3}}
\newcommand{\rcite}[1]{\hyperlink{cite.#1}{~\cite{#1}}}
\newcommand{\eqqref}[1]{\hyperref[#1]{Eq.~\eqref{#1}}}
\newcommand{\tabref}[1]{\hyperref[#1]{~Tab.~\ref{#1}}}
\newcommand{\chid}[2]{\chi_{\text{#1}}^{#2}}

\def\@#1{_{\rm #1}}

\begin{document}

\preprint{APS/123-QED}

\title{Angle-Invariant Scattering in Metasurfaces}

\author{Mustafa Yücel}%
\email{mustafa.yucel@epfl.ch}
\author{Francisco S. Cuesta}%
\author{Karim Achouri}%
 \email{karim.achouri@epfl.ch}
\affiliation{%
Institute of Electrical and Microengineering, École Polytechnique Fédérale de Lausanne (EPFL), Laboratory for Advanced Electromagnetics and Photonics, Lausanne, Switzerland\\}%

\date{\today}

\begin{abstract}

Metasurfaces are efficient and versatile electromagnetic structures that have already enabled the implementation of a wide range of microwave and photonic wave shaping applications. Despite the extensive research into metasurfaces, a rigorous and comprehensive understanding of their angular dispersion remains vastly under-explored. Here, we use the generalized sheet transition conditions (GSTCs) to model and analyze the angular dispersive properties of metasurfaces. Based on this theoretical framework, we demonstrate that a metasurface may exhibit either partial or complete co- and cross-polarized transmission and reflection coefficients that are angle-invariant, meaning that their amplitude, phase, or both remain unchanged with varying incidence angles. We show that these angle-independent responses exist only when specific conditions, given in terms of the metasurface effective susceptibilities, are met. Using the GSTCs formalism, we derive several of these conditions and illustrate their scattering properties. Among other findings, this analysis reveals that, contrary to common assumptions, nonlocality (spatial dispersion) does not only increase the angular dispersion of a metasurface, but may also be used to achieve complete angle-invariant scattering. Additionally, this work demonstrates that fully efficient extrinsic chirality is possible with a pseudochiral metasurface in a partially angle-invariant fashion. We expect our work to provide a general strategy for eliminating, or at least reducing, angular-dependent scattering responses of metasurfaces, which may prove instrumental for applications that are highly sensitive to the detrimental effects of angular dispersion.

\end{abstract}

\maketitle


\section{\label{sec:intro}Introduction  \protect }

Metamaterials are artificially engineered structures designed to manipulate the propagation of electromagnetic waves in forms that transcend the limitations of natural materials. Today, a ubiquitous form of these man-made materials is metasurfaces -- thin two-dimensional arrays of engineered scattering particles -- that came to prominence due to their ease of fabrication and limited losses compared to volumetric metamaterials. Over the past decade, metasurfaces have garnered significant attention for their unique capabilities, including achieving negative or near-zero effective material parameters and enabling a wide range of optical phenomena and applications.\rcite{chen2016review,hu2021review,kuznetsov2024roadmap,schulz2024}. Despite extensive research, the field of metasurfaces continues to expand rapidly, driven by theoretical and technological advancements that open new possibilities and enhance their capabilities across diverse fields, from microwave systems to quantum computing.

An essential feature of metasurfaces is their ability to control light properties such as direction of propagation, frequency, polarization, phase, and amplitude. This capability is leveraged in a broad spectrum of applications including computational imaging\rcite{momeni2019,zheng2021metasurface}, augmented and virtual reality\rcite{liu2023metasurface}, biosensing\rcite{patel2020graphene}, nonlinear optics\rcite{li2017nonlinear}, and quantum and topological optics\rcite{kan2020directional}.
\begin{figure}[H]
    \centering
    \includegraphics[width=\linewidth]{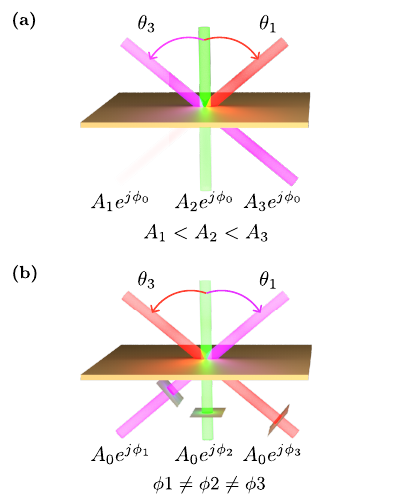}
    \caption{\textbf{Transmission characteristics for different angles of incidence illustrated with different colors. (a)} Phase remains angle-invariant while amplitude varies.\textbf{(b)} Amplitude is angle-invariant, but the phase varies.}
    \label{intro_fig}
\end{figure}
Despite the wide array of properties that metasurfaces can handle, the characteristic of \emph{angle-invariance scattering} remains vastly unexplored and only a few studies have so far discussed this topic. One of the first description of angle-invariant reflection and transmission amplitude in isotropic metasurfaces was provided in\rcite{holloway2009a}. Then, the concept of angularly-independent Huygens' metasurfaces was introduced in\rcite{radi2015a}, where it was shown that uniaxial structures may exhibit angle-independent full transmission with arbitrary phase shifts. On a similar note, the concept of all-angle transparency in anisotropic metasurfaces was further extended both theoretically and experimentally in~\cite{shaham2024}. Other studies have discussed the problem of angular dispersion in metasurfaces and demonstrated that angle-insensitive absorbers may be implemented experimentally, as discussed in\rcite{qiu2018angular,zhang2020controlling}. Finally, the concept of purely nonlocal (with only bianisotropic responses) angle-invariant reflectance was introduced in~\cite{lavigne2019,achouri2021,achouri2020a}. Despite these early efforts, a significant gap remains in understanding the phenomenon of angular invariance in the broader context of dipolar bianisotropic metasurfaces. 

This paper aims to address this gap in understanding by analyzing the angular characteristics of metasurfaces through their unit-cell effective susceptibilities and geometries using a modeling framework based on the General Sheet Transition Conditions (GSTCs)~\cite{kuester2003,achouri2021}.
We demonstrate that a metasurface can exhibit angular scattering invariance in either phase, amplitude or both under specific conditions, as illustrated in\figref{intro_fig}{}. Additionally, we explore the design, shape and spatial symmetries that the unit cells must adopt to achieve a variety of different angular invariance conditions. Our theoretical approach can be applied to both transverse electric (TE) and transverse magnetic (TM) polarizations as well as for circularly polarized beams. It is also not restricted to a single plane of incidence but rather covers the entire range of propagation angles within the visible light cone. 
To provide a general perspective on the problem of angle-invariant scattering, we start by discussing angle scattering invariance in the case of isolated particles in~\secref{sec:isolated_particle}. Then, we study the angle scattering invariance in metasurfaces using the GSTCs modeling framework in~\secref{sec:metasurface}. In~\subsecref{subsec:gstc_theory}, we review the theoretical background behind the GSTCs and explain how they may be applied to model the scattering response of a metasurface. As a proof of concept, we show meta-devices that exhibit partial or complete angle-invariant phase and/or amplitude in~\subsecrange{subsec:phase_inv}{subsec:amplitude_inv}{subsec:total_inv}. Our findings can help understand and manipulate the angular response of metasurfaces, which may significantly expand our capabilities of designing metasurfaces-based functional devices.
\section{\label{sec:isolated_particle} Angle Scattering Invariance in Isolated Particles}
\begin{figure}[h]
    \centering
    \includegraphics[width=\linewidth]{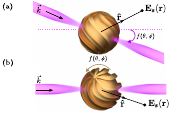}
    \caption{\textbf{Scenarios for evaluating angular scattering invariance}: \textbf{(a)} the incident beam rotates around a fixed particle, and \textbf{(b)} the particle itself rotates.}
    \label{fig:particle_rot}
\end{figure}

Before delving into the scattering properties of a metasurface, we shall first investigate the case of angular invariance in an isolated scatterer. Let us imagine a randomly shaped deeply subwavelength scatterer that is excited by an incoming beam propagating with the wavevector $\ve{k}$. To study the angular response of this scatterer, we can either consider that the beam impinges on a fixed scatterer with different angles of incidence, as depicted in\figref{fig:particle_rot}{a}, or a fixed beam illuminating a rotating particle, as in\figref{fig:particle_rot}{b}. At first glance, one may a priori think that the two cases depicted in\figref{fig:particle_rot}{} are identical, however, this is not generally true.

To investigate the differences between the two cases depicted in\figref{fig:particle_rot}{} and determine under which conditions the particle exhibits angle-invariant scattering, we have to analyze its electromagnetic behavior. To do so, we next consider that the particle is deeply subwavelength and thus express its electromagnetic scattering response in terms of its electric and magnetic dipole moments, $\ve{p}$ and $\ve{m}$, respectively. The corresponding electric field scattered by this particle in the far region (assumed to be in vacuum) is then given by~\cite{jackson2009}
\begin{equation}
    \ve{E}_\text{s} = \frac{G\@{0}}{\epsilon_0}\left[\ve{\hat{r}}\times(\ve{p}\times\ve{\hat{r}})+\frac{1}{c_0}\ve{m}\times\ve{\hat{r}}\right],    \label{eq:scattered_electric_field_particle}
\end{equation}
where $G\@{0} = \frac{k^2}{4\pi}\frac{e^{jkr}}{r}$ is the scalar Green function, $\ve{\hat{r}}$ is a unit vector in spherical coordinates pointing in the direction between the particle and the position of the scattered field $\ve{E}_\text{s}$, $r$ is the distance between the object and the point of observation, and $k$ is the wavenumber. In the general situation, where the particle does not exhibit any spatial symmetry, a bianisotropic formalism is required to model its response~\cite{poleva2023}. In this case, its dipole moments may be expressed in terms of the incident field as\rcite{tretyakov2003}
\begin{equation}
\label{eq_pm}
\begin{bmatrix}
    \ve{p}\\
    \ve{m}
\end{bmatrix}=
\begin{bmatrix}
    \epsilon_0\te{\alpha}\@{ee} & \frac{1}{c_0} \te{\alpha}\@{em}\\
    \frac{1}{\eta_0}\te{\alpha}\@{me} &  \te{\alpha}\@{mm}
\end{bmatrix}\cdot
\begin{bmatrix}
    \ve{E}\@{i}\\
    \ve{H}\@{i}
\end{bmatrix},
\end{equation}
where the $\te{\alpha}$ terms correspond to the particle polarizability tensors.
\newline

\begin{figure}[h]
    \centering
    \includegraphics[width=\linewidth]{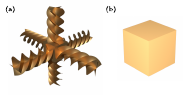}
    \caption{\textbf{Bi-isotropic and (mono)-isotropic isolated particles. (a)} Helices induce magneto-electric coupling by breaking reflection symmetry, this is characteristic of bi-isotropic materials. \textbf{(b)} In a simple cube exhibiting inversion symmetry, magneto-electric coupling vanishes and the material response is (mono)-isotropic.}
    \label{fig_bi_isotropic_and_isotropic}
\end{figure}

Since we are interested in angular scattering invariance, we now look for the conditions that apply on the system in~\eqqref{eq_pm} such that the particle is invariant under a rotation operation.
Meaning that, for a fixed illumination direction and polarization state, the field scattered by the particle remains identical irrespectively of the rotation of the object, as suggested in\figref{fig:particle_rot}{b}. 
This may be achieved by considering that if the particle undergoes a spatial transformation (e.g., rotation, reflection, etc) given by the operator $\te{\Lambda}$ and is assumed to be invariant under that transformation, then it must necessarily satisfy the following invariance conditions~\cite{arnaut1997,achouri2023a}
\begin{subequations}
\label{eq_alpha_eemm}
\begin{align}
    \te{\alpha}\@{ee/mm} &= \te{\Lambda} \cdot \te{\alpha}\@{ee/mm} \cdot \te{\Lambda}^T \label{eq:alpha_ee} \\
    \te{\alpha}\@{em/me} &= \det\left(\te{\Lambda}\right) \te{\Lambda} \cdot \te{\alpha}\@{em/me} \cdot \te{\Lambda}^T,
    \label{eq:alpha_mm}
\end{align}
\end{subequations}
where $\te{\Lambda}^T$ is the transpose of the operator $\te{\Lambda}$, which is assumed to be an orthogonal matrix. Such rotation invariance may be imposed on an arbitrary scatterer by forcing its polarizability tensors to be invariant under a transformation defined by $\te{\Lambda} = R_i(\theta)$, where $R_i(\theta)$ is a rotation matrix around the axis $i \in \{x,y,z\}$ by the angle $\theta$~\cite{achouri2023a}. Applying such transformations along at least two different axes and for $\theta=2\pi/N$ with $N\geq 3$ to the polarizability tensors in~\eqqref{eq_pm} so that they satisfy~\eqqref{eq_alpha_eemm}, yields
\begin{subequations}
\label{eq_biiso}
\begin{align}
\te{\alpha}\@{ee} &= \alpha\@{ee} \te{I},\quad\quad \te{\alpha}\@{mm} = \alpha\@{mm} \te{I}, \\
\te{\alpha}\@{em} &= \alpha\@{em} \te{I},\quad\quad \te{\alpha}\@{me} = \alpha\@{me} \te{I},
\end{align}
\end{subequations}
where $\te{I}$ is the identity tensor. These polarizability tensors are isotropic and thus satisfy~\eqqref{eq_alpha_eemm} for any rotation. Note that a scatterer described by the tensors in~\eqqref{eq_biiso} is bi-isotropic due to the presence of the magneto-electric coupling tensors $\te{\alpha}\@{me}$, $\te{\alpha}\@{em}$. An example of a scatterer that satisfies this case for $N=4$ is illustrated in\figref{fig_bi_isotropic_and_isotropic}{(a)}.

Note that further imposing reflection symmetries $\sigma_i$, with $i\in\{x,y,z\}$, reduces~\eqqref{eq_biiso} to an isotropic system for which $\te{\alpha}\@{me}=\te{\alpha}\@{em} = 0$ and $\te{\alpha}\@{ee} =\alpha\@{ee} \te{I}, \te{\alpha}\@{mm} = \alpha\@{mm} \te{I}$~\cite{achouri2023a}. A simple geometry that corresponds to this case is that of a cube, as illustrated in\figref{fig_bi_isotropic_and_isotropic}{(b)}, which is both rotationally invariant (for $N=4$) and reflection invariant.

Now that we understand that a bi-isotropic particle with the polarizabilities as in~\eqqref{eq_biiso} is rotationally invariant and would correspond to the situation illustrated in\figref{fig_bi_isotropic_and_isotropic}{(b)}, we want to verify whether it would also satisfy the situation illustrated in\figref{fig_bi_isotropic_and_isotropic}{(a)}. For this purpose, we substitue~\eqqref{eq_pm}, along with the polarizabilities in~\eqqref{eq_biiso}, into~\eqqref{eq:scattered_electric_field_particle} and re-express the electric field scattered by the particle in terms of the incident electric field (see \textbf{Appendix~A} in the Supplementary Information -- SI) as
\begin{equation} 
\ve{E}_\text{s}= G\@{0}\te{G}(\ve{k})\cdot\ve{E}\@{i},
\end{equation}
where $\te{G}(\ve{k})$ is a dyadic Green function that may be split in two separate contributions as
\begin{equation}
    \te{G}(\ve{k}) = \te{G}_1(\ve{k}) + \te{G}_2(\ve{k}),
\end{equation}
with
\begin{subequations}
\label{eq_G12}
\begin{align}
\te{G}_1(\ve{k})&=\alpha\@{ee}\te{I}_\parallel-\alpha\@{mm}\left[\ve{\hat{k}}\ve{\hat{r}}-\left(\ve{\hat{r}}\cdot\ve{\hat{k}}\right)\te{I}\right],\\
\te{G}_2(\ve{k}) &=\left[\alpha\@{em}\te{I}_\parallel+\alpha\@{me}\left[\ve{\hat{k}}\ve{\hat{r}}-\left(\ve{\hat{r}}\cdot\ve{\hat{k}}\right)\te{I}\right]\right]\cdot\te{k},
\end{align}
\end{subequations}
where $\te{I}_\parallel = \te{I} - \ve{\hat{r}}\ve{\hat{r}}$ and $\te{k} = \ve{\hat{k}}\times\te{I}$. Upon inspection, the functions $\te{G}_1(\ve{k})$ and $\te{G}_2(\ve{k})$ are both rotation invariant, as explained in the \textbf{Appendix~A} of the SI and which is consistent with the fact that they are defined using rotationally invariant polarizabilities. However, and maybe more surprisingly, they are also both dependent on the wavevector $\ve{k}$. Meaning that a particle that satisfies the circumstances in\figref{fig:particle_rot}{(b)} generally does not satisfy the ones of\figref{fig:particle_rot}{(a)}. While this might a priori seem counter-intuitive, it may be understood as a consequence of the fact that one needs to consider the symmetries of the entire system (particle + illumination) and not just the spatial symmetric of the particle alone. Another way of understanding this is by considering that a change in the direction of propagation of the incident wave leads to a phase shift between its electric and magnetic field components, which, in turn, changes the overall scattering response of the particle. Therefore, a (bi)-isotropic dipolar particle is generally both rotationally invariant and $\ve{k}$-dependent, even in the absence of the magneto-electric constants (i.e., $\alpha_\text{em} = \alpha_\text{me} = 0$). The only situation where a particle would be both rotationally and $\ve
{k}$ invariant is when its scattering response may be fully modeled with only $\alpha_\text{ee}$ leading to a scattered field given by
\begin{equation}
    \ve{E}_\text{s}= G\@{0}\alpha\@{ee}\left( \te{I} - \ve{\hat{r}}\ve{\hat{r}}\right)\cdot\ve{E}\@{i}.
\end{equation}
It remains interesting to consider the special cases where the observation direction is collinear with the $\ve{k}$-vector such that $\ve{\hat{r}}= \pm \ve{\hat{k}} =\pm \ve{k}_\text{i}/|\ve{k}_\text{i}|$. This reduces~\eqqref{eq_G12} to
\begin{subequations}
\begin{align}
\te{G}_1(\ve{k})&=\left(\alpha\@{ee}\pm\alpha\@{mm}\right)\te{I},\\
\te{G}_2(\ve{k}) &=\left(\alpha\@{em}\mp\alpha\@{me}\right)\te{k}.
\end{align}
\end{subequations}
We see that when restricting ourselves to the forward and backward scattering directions with respect to the propagation direction of the incident wave, the function $\te{G}_1(\ve{k})$ loses it dependency on the $\ve{k}$-vector. Upon inspection of $\te{G}_1(\ve{k})$, we directly retrieve the dipolar Kerker conditions $\alpha\@{ee}=-\alpha\@{mm}$ and $\alpha\@{ee}=\alpha\@{mm}$ that correspond to a nullified forward and backward scattering, respectively~\cite{kerker_1983,kerker_2013}. On the other hand, $\te{G}_2(\ve{k})$ remains generally $\ve{k}$-dependent when $\{\alpha\@{em},\alpha\@{me}\}\neq 0$. At the exception of the two special cases where $\alpha\@{em}=\alpha\@{me}$, when $\ve{\hat{r}}=\ve{\hat{k}}$, and $\alpha\@{em}=-\alpha\@{me}$, when $\ve{\hat{r}}=-\ve{\hat{k}}$, leading to $\te{G}_2(\ve{k}) = 0$. In the former case, the particle is nonreciprocal and may be classified as a Tellegen particle, while in the latter case, the particle is reciprocal and may be classified as a Pasteur particle~\cite{tretyakov2001}. We therefore conclude that a reciprocal bi-isotropic particle is always rotationally invariant but is only $\ve{k}$-invariant when observing the field scattered in the backward direction ($\ve{\hat{r}}=-\ve{\hat{k}}$).
\section{\label{sec:metasurface} Angle-Invariant Scattering in Metasurfaces}
\begin{figure}[h]
    \centering
    \includegraphics[width=\linewidth]{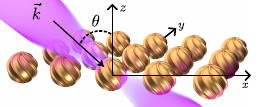}
    \caption{\textbf{Bianostropic metasurface being illuminated at oblique incidence}. The period of the array is small enough compared to the wavelength so that only the zeroth-order diffracted waves are allowed to propagate. In this case, the scattering response of the metasurface may be modeled by homogeneous effective material parameters.}
    \label{fig:fig_metasurface_particle}
\end{figure}

%
As mentioned earlier, a metasurface is an array of particles that alters the properties of an incident beam, as represented in\figref{fig:fig_metasurface_particle}. Unlike isolated particles, analyzing angular invariance in metasurfaces requires a different approach. We use the GSTCs to develop a mathematical model and analyze the metasurface's response.

Intuitively, one may a priori think that the ability to achieve angle-invariant scattering in a metasurface does not significantly differ from what we have seen for an isolated particle. However, we will see that this not the case, and that a metasurface presents a rich diversity of conditions leading to angle-invariant scattering.

\subsection{\label{subsec:gstc_theory} Theoretical Background}
To model the electromagnetic response of a metasurface and investigate its angular scattering behavior, we use the GSTCs which are given by~\cite{achouri2021}
\begin{subequations}
\label{eq_GSTC}
\begin{align}
\hat{z} \times \Delta \ve{H} &= j \omega \ve{P}_\parallel - \hat{z} \times \nabla_\parallel M_z, \\
\hat{z} \times \Delta \ve{E} &= -j \omega \mu_0 \ve{M}_\parallel - \hat{z} \times \nabla_\parallel \left(\frac{P_z}{\epsilon_0}\right),
\end{align}
\end{subequations}
where $\Delta \ve{H} = \ve{H}^{z=0^+}_\parallel-\ve{H}^{z=0^-}_\parallel$ and $\Delta \ve{E} = \ve{E}^{z=0^+}_\parallel-\ve{E}^{z=0^-}_\parallel $ and the subscript $\parallel$ indicates tangential vector components with respect to the $xy$-plane. In the general case of bianisotropic metasurface, the constitutive relations associated with~\eqqref{eq_GSTC} are defined by
\begin{subequations}
\label{eq_PM}
\begin{align}
\ve{P} &= \epsilon_0 \te{\chi}\@{ee} \cdot \ve{E}\@{av} + \frac{1}{c_0} \te{\chi}\@{em} \cdot \ve{H}\@{av}, \\
\ve{M} &= \te{\chi}\@{mm} \cdot \ve{H}\@{av} + \frac{1}{\eta_0} \te{\chi}\@{me} \cdot \ve{E}\@{av},
\end{align}
\end{subequations}
where $\epsilon_0$ and $\mu_0$ are the dielectric and the permeability constants in vacuum, respectively. $\ve{E}\@{av}$ and $\ve{H}\@{av}$ are the averaged electric and magnetic fields at the interface, respectively. $\te{\chi}\@{ee}$, $\te{\chi}\@{mm}$, $\te{\chi}\@{em}$ and $\te{\chi}\@{me}$ are the electric, magnetic, magnetic-to-electric and electric-to-magnetic susceptibility tensors, respectively~\cite{achouri2021}.

In the following discussion, the metasurfaces will exhibit arbitrary electric and magnetic dipoles oriented along a chosen axis. We will however, restrict our attention to the case where the metasurface is uniform and homogeneous such that its susceptibilities are not spatially varying. This implies that the scattering particles forming the metasurface are repeated with a subwavelength period (smaller than $\lambda/2$, with $\lambda$ as the illuminating source wavelength). By selecting the shape of the particles, specific susceptibilities can be achieved by adhering to the spatial symmetry principles described in\rcite{achouri2023a}. We also consider that the metasurfaces that we will discuss are reciprocal, which requires the following conditions to be satisfied~\cite{achouri2021}
\begin{equation}
\label{eq_recip}
    \te{\chi}_\text{ee} =   \te{\chi}_\text{ee}^{T}, \quad \te{\chi}_\text{mm} =   \te{\chi}_\text{mm}^{T}, \quad \te{\chi}_\text{em} =  -\te{\chi}_\text{me}^{T},
\end{equation}
%
If we also impose that these metasurfaces are passive and lossless, then they must additionally satisfy the conditions~\cite{achouri2021}
\begin{equation}
\label{eq_LP}
    \te{\chi}_\text{ee} =   \te{\chi}_\text{ee}^\dagger, \quad \te{\chi}_\text{mm} =   \te{\chi}_\text{mm}^\dagger, \quad \te{\chi}_\text{em} = \te{\chi}_\text{me}^\dagger,
\end{equation}
where $\dagger$ corresponds to the conjugate transpose operator. The combination of~\eqqref{eq_recip} and~\eqqref{eq_LP} leads to the implication that a lossless, passive and reciprocal metasurface must necessarily satisfy
\begin{equation}
\label{eq_recipLP}
    \Im(\te{\chi}_\text{ee}) = \Im(\te{\chi}_\text{mm}) =\Re(\te{\chi}_\text{em}) = \Re(\te{\chi}_\text{me}) = 0,
\end{equation}
where $\Im(.)$ and $\Re(.)$ describe the imaginary and the real parts of a complex number, respectively.
\subsection{Diagonal Anisotropic Metasurfaces}
Before diving into the general case of bianisotropic structures, we start by considering the simpler case of a diagonal anisotropic metasurface, for which the magnetic-to-electric and electric-to-magnetic susceptibilities are zero, i.e., $\te{\chi}\@{me} = \te{\chi}\@{em} = 0$, as well as the off-diagonal elements of $\te{\chi}\@{mm}$ and $\te{\chi}\@{ee}$. 

For conciseness, we shall consider the case of TM-polarized wave propagating in the $xz$-plane. In this case, a metasurface can only respond to such an excitation via the susceptibilities $\chi_\text{ee}^{xx}$, $\chi_\text{ee}^{zz}$ and $\chi_\text{mm}^{yy}$. Note that if the TE polarization is considered instead, then these susceptibilities should be replaced by $\chi_\text{mm}^{xx}$, $\chi_\text{mm}^{zz}$ and $\chi_\text{ee}^{yy}$, respectively.
We now assume that the metasurface is illuminated by a plane wave that is specularly reflected and transmitted. The corresponding co-polarized transmission and reflection equations are found by inserting the mathematical definition of these plane waves into~\eqqref{eq_GSTC} along with~\eqqref{eq_PM}, as detailed in \textbf{Appendices B and C1} of the SI, as well as in~\cite{achouri2021,achouri2020a}. The resulting co-polarized reflection and transmission coefficients are
\begin{subequations}
\label{eq_coTR_diagAnis}
\begin{align}
R_{\text{co}}(\theta) &= \frac{2 \left(k_{x}^2 \chi_{\text{ee}}^{{zz}} + k^2 \chi_{\text{mm}}^{{yy}} - k_{z}^2 \chi_{\text{ee}}^{{xx}}\right)}{\left(k_{z} \chi_{\text{ee}}^{{xx}} - 2j\right) \left[2 k_{z} + j \left(k_{x}^2 \chi_{\text{ee}}^{{zz}} + k^2 \chi_{\text{mm}}^{{yy}}\right)\right]}, \\
T_{\text{co}}(\theta) &= -\frac{j k_{z} \left(4 + k_{x}^2 \chi_{\text{ee}}^{{xx}} \chi_{\text{ee}}^{{zz}} + k^2 \chi_{\text{ee}}^{{xx}} \chi_{\text{mm}}^{{yy}}\right)}{\left(k_{z} \chi_{\text{ee}}^{{xx}} - 2j\right) \left[2 k_{z} + j \left(k_{x}^2 \chi_{\text{ee}}^{{zz}} + k^2 \chi_{\text{mm}}^{{yy}}\right)\right]},
\end{align}
\end{subequations}
where the components of the wavevector are generally given by $k_x=k \sin\theta\cos\phi$, $k_y=k \cos\theta\sin\phi$ and \mbox{$k_z=k \cos\theta$}. In the case of~\eqqref{eq_coTR_diagAnis}, we have that $\phi=0$ since we are only considering the metasurface scattering response within the $xz$-plane. Note that for a diagonal anisotropic metasurface, the cross-polarized scattering coefficients are zero. 

We are now interested in determining the conditions on the susceptibilities for which these scattering coefficients become angle-invariant. Remembering that the susceptibilities should not be functions of the angle of incidence, it turns out that it is impossible to achieve full (both amplitude and phase) angle invariance with such coefficients. This stems from the fact that it is impossible to find susceptibilities such that the angular dependence provided by the wavevector components, $k_x$ and $k_z$, completely cancels out in the expressions of the scattering parameters. However, if we concentrate our attention only on the phase and amplitude of these coefficients separately, then it becomes possible to remove their angular dependence. This is the topic of the next two subsections, respectively.
\begin{figure}[h]
    \centering
    \includegraphics[width=\linewidth]{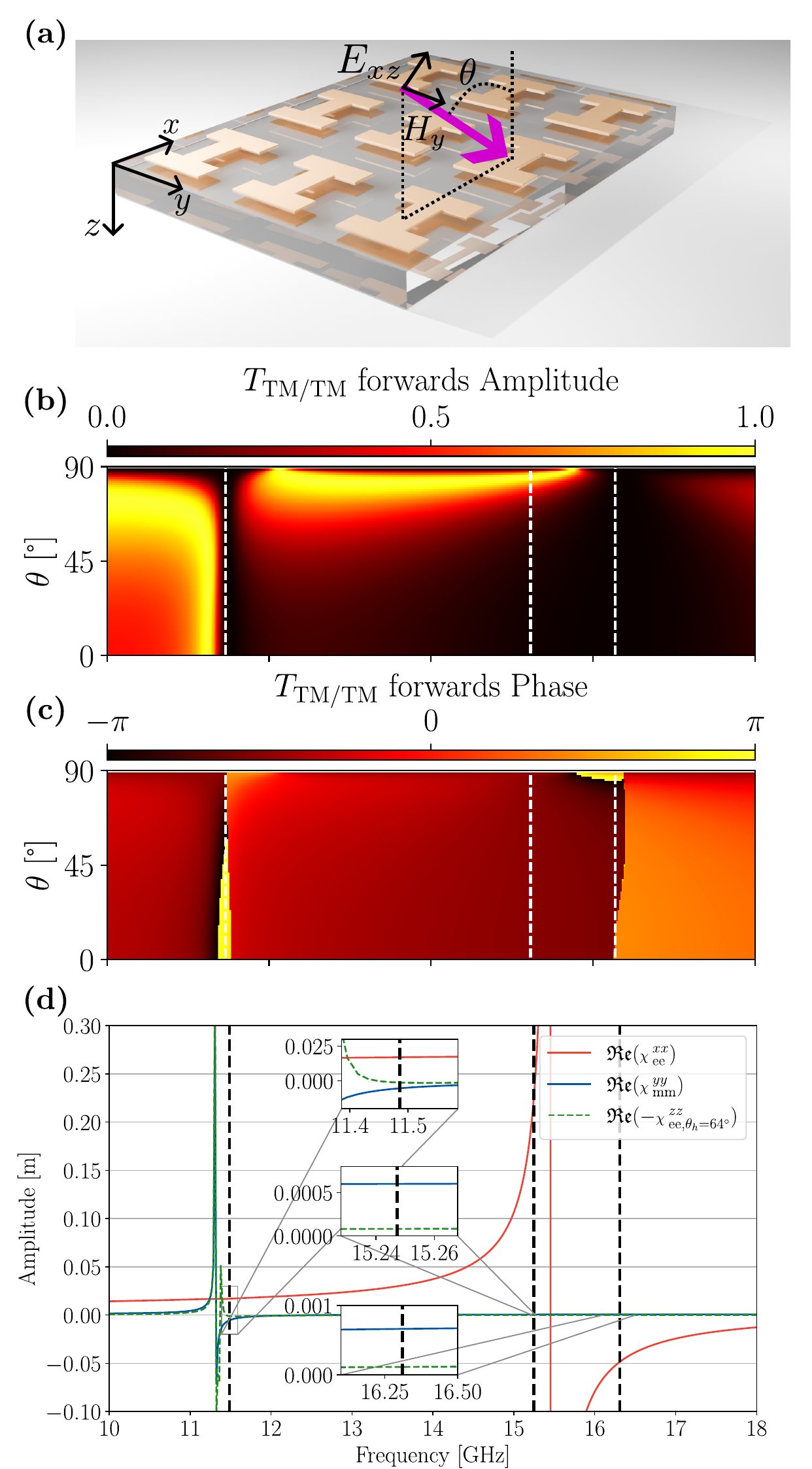}
    \caption{\textbf{Dog bone structure simulation in microwave.} \textbf{(a)} A TM mode impinges on the dog bone array. \textbf{(b)} Simulated co-polarized transmission amplitude and \textbf{(c)} phase for different angle and frequency ranging from $0^\circ$ to $90^\circ$ and $10$ GHz to $18$ GHz. \textbf{(d)} Retracted susceptibilities from the simulation and GSTC model. }
    \label{fig:spectrum_and_susceptibility}
\end{figure}
\subsubsection{\label{subsec:phase_inv} Angle-Invariant Phase}
In order to investigate the phase response of the metasurface scattering parameters, we split the coefficients in~\eqqref{eq_coTR_diagAnis} into their real and imaginary parts and express their corresponding phase as
\begin{subequations}
\begin{align}
    \angle R_\text{co} &= \text{tan}^{-1}\left(\frac{\Im(R_\text{co}(\theta))}{\Re(R_\text{co}(\theta))}\right),\\
    \angle T_\text{co} &= \text{tan}^{-1}\left(\frac{\Im(T_\text{co}(\theta))}{\Re(T_\text{co}(\theta))}\right).
\end{align}
\end{subequations}
Knowing that the coefficients in~\eqqref{eq_coTR_diagAnis} are not angle-invariant, it is clear that their real and imaginary parts \emph{cannot be simultaneously} angle-invariant either. Additionally, it may be shown that their ratio is also not angle-invariant. This leaves only two possibilities to achieve an angle-invariant phase. Namely, either $\Im(R_\text{co}(\theta)) = 0$ and  $\Re(R_\text{co}(\theta)) \neq 0$ or the opposite, i.e., $\Im(R_\text{co}(\theta)) \neq 0$ and  $\Re(R_\text{co}(\theta)) = 0$, and similarly for the transmission coefficient. This directly implies that the phase of $R_\text{co}$ and $T_\text{co}$ can only take values of $n\pi/2$ where $n\in \mathbb{N}$.

Accordingly, to obtain conditions on the susceptibilities, we now solve the equations
\begin{subequations}
\label{eq_condphaseinv}
\begin{align}
\Re(R_\text{co}(\theta)) & = 0, & \quad \Im(R_\text{co}(\theta)) & = 0,\label{eq_condphaseinvR} \\
\Re(T_\text{co}(\theta)) & = 0, & \quad \Im(T_\text{co}(\theta)) & = 0,\label{eq_condphaseinvT}
\end{align}
\end{subequations}
for the susceptibilities, this leads to two solutions per equation [as shown in the SI~\textbf{(C9)-(C12)}], altough we are primarily interested in only few of them. It follows that the equations $\Im(R_\text{co}(\theta)) = 0$ and $\Re(T_\text{co}(\theta)) = 0$ are both satisfied by
\begin{equation}
\chi_{\text{mm}}^{{yy}}(\theta) = \frac{4 - k_x^2 \chi_{\text{ee}}^{{xx}}\chi_{\text{ee}}^{{zz}}}{\chi_{\text{ee}}^{{xx}}k^2}. 
\end{equation}
It is clear that this equation is angle dependent due to the presence of $k_x^2$. However, to achieve a fully angle-independent solution, we can set $\chi_{\text{ee}}^{zz} = 0$, so that the angular dependence of the solution is eliminated, resulting in the fully angle-independent solution prescribed by
\begin{subequations}
\label{angle_indep_cond1}
\begin{align}
    \chi_\text{ee}^{zz} &= 0, \\
    \chi_\text{mm}^{yy} \Big|_{\chi_{\text{ee}}^{zz} = 0}&= \frac{4}{\chi_\text{ee}^{xx} k^2}.
    \end{align}    
\end{subequations}
If these conditions are satisfied, then the corresponding metasurface would induce a constant phase shift due to the transmission coefficient \( T_\text{co}(\theta) \) being purely imaginary and the reflection coefficient \( R_\text{co}(\theta) \) being purely real, as presented in the first row of\tabref{table_summary_diagonal_anis}. To achieve a scenario where the susceptibility \( \chi_{\text{ee}}^{zz} = 0 \), one possible approach is to design a structure with minimal thickness, limiting its polarization response to tangential components. Alternatively, leveraging frequency dispersion to meet this criterion provides another viable strategy. For instance, it is possible to design a structure harboring several Lorentzian type resonances and exploit a limited frequency range where $\chi_{\text{ee}}^{zz}$ approaches zero somewhere in between these resonances.

To demonstrate that it is indeed possible to achieve the conditions in~\eqqref{angle_indep_cond1}, we simulate the double dog-bone structure shown in\figref{fig:spectrum_and_susceptibility}{(a)}. The structure has been selected by applying the symmetry considerations described in~\cite{achouri2023a} to obtain a biaxial anisotropic structure whose physical details are given in~\text{\color{blue}Fig. S1} of the SI. Figures~\ref{fig:spectrum_and_susceptibility}{(b-c)} show the full-wave simulated TM-polarized co-transmission coefficient amplitude and phase, respectively. We can observe that in \figref{fig:spectrum_and_susceptibility}{(b-c)}, at the position of the second dashed line, the transmission phase does not change with the angle of incidence, whereas the transmission amplitude varies significantly. This constant angular phase occurring at 12.25~GHz corresponds to a phase shift of $-\pi/2$, which implies that the transmission is negative and imaginary, i.e., $T_\text{co}(\theta) = -j|\kappa(\theta)|$ where $\kappa(\theta) \in \mathbb{C}$. This suggests that it may correspond to the transmission response shown in the first row in\tabref{table_summary_diagonal_anis}. To further verify that we are close to this condition, we compute the effective susceptibilities $\chid{ee}{xx}$, $\chid{ee}{zz}$ and $\chid{mm}{yy}$ of this metasurface following the procedure described in the SI. The resulting retrieved susceptibilities are plotted in \figref{fig:spectrum_and_susceptibility}{(d)}, where it can be observed in the insets that the conditions in~\eqqref{angle_indep_cond1} are almost met.
\newline 

Interestingly, the conditions in~\eqqref{angle_indep_cond1} could be linked to the Kerker condition~\cite{kuznetsov2016optically,achouri2021}, for which $R\@{co}=0$ at normal incidence. This is achieved by imposing that the electric susceptibility, $\chi_\text{ee}^{xx}$, is equal to the magnetic susceptibility, $\chi_\text{mm}^{yy}$, such that
\begin{align}
\label{eq_kerker}
    \chi_\text{mm}^{yy} = \chi_\text{ee}^{xx} = \chi_\text{k}, 
\end{align}
where $\chi_\text{k}$ is constrained to $\chi_\text{k} = \pm 2/k$ to satisfy~\eqqref{angle_indep_cond1}. Notice that the combination of this Kerker condition with the conditions in~\eqqref{angle_indep_cond1} leads to a compromise. It limits the achievable phase shift of the transmission coefficient to $\pm \pi/2$ and that of the reflection coefficient to 0. This means that forcing the transmission amplitude to be unity at normal incidence necessarily implies that the angle-invariant transmission phase shift is limited to a binary value.  \newline

We now look at the case where $\Re(R_\text{co}(\theta)) = 0$ and $\Im(T_\text{co}(\theta)) = 0$. Considering~\eqqref{eq_condphaseinv}, the new condition leading to an angle-invariant phase is
\begin{align}
    \chi_{\text{mm}}^{{yy}}(\theta) = -\frac{k_z^2 \chi_{\text{ee}}^{{xx}} + k_x^2 \chi_{\text{ee}}^{{zz}}}{k^2}.
\end{align}
It is again visible that the solution has an angle dependency due to the presence of $k_x^2$ and $k_z^2$. To achieve a fully angle-independent solution, we set $\chi_\text{ee}^{xx} = \chi_\text{ee}^{zz}$ to obtain $k_z^2 + k_x^2 = k^2$ in the numerator, which cancels out the angular dependency. We thus have
\begin{subequations}
\begin{align}
    \chi_\text{ee}^{zz} & = \chi_\text{ee}^{xx}, \\
    \chi_\text{mm}^{yy} \Big|_{\chi_\text{ee}^{zz} = \chi_\text{ee}^{xx}} &= -\chi_\text{ee}^{xx}.
\end{align}
\label{angle_indep_cond2}
\end{subequations}
With these conditions, the angular dependency is removed from the solution resulting in a purely real transmission coefficient, $T_\text{co}(\theta)$, and a fully imaginary reflection coefficient, $R_\text{co}(\theta)$. The resulting expressions for these coefficients are given in the second row of\tabref{table_summary_diagonal_anis}.

\begin{table}[h!]
\caption{\textbf{Summary table for diagonal-anisotropic material}. Note the relevent symmetry between the conditions in the first and third rows, as well as the second and fourth rows.}
\renewcommand{\arraystretch}{2} 
\begin{tabular}{c|c|c|c|}
\cline{2-4}
                                        & Conditions & Implications & Expression \\ \hline
\multicolumn{1}{|l|}{\multirow{2}{*}{}} & 
\makebox[2cm][c]{
    \begin{minipage}{2cm}
        \vspace{0.5em} 
        \begin{align*}
            \chi_\text{ee}^{zz} & = 0 \\
            \chi_\text{mm}^{yy}\chi_\text{ee}^{xx} & = \frac{4}{k^2}
        \end{align*}
        \vspace{0.5em} 
    \end{minipage}
} & 
\makebox[2cm][c]{
    \begin{minipage}{2cm}
        \vspace{0.5em} 
        \begin{align*}
            T_\text{co}(\theta) & \in \mathbb{I} \\
            R_\text{co}(\theta) & \in \mathbb{R}
        \end{align*}
        \vspace{0.5em} 
    \end{minipage}
} &
\makebox[3.5cm][c]{
    \begin{minipage}{2cm}
        \vspace{0.5em} 
        \begin{align*}
            T_\text{co}(\theta) & = -\frac{4jk_\text{z}\chi_\text{ee}^{xx}}{4 + k_\text{z}^2\chi_\text{ee}^{xx2}} \\
            R_\text{co}(\theta) & = \frac{4 - k_\text{z}^2\chi_\text{ee}^{{xx}^2}}{4 + k_\text{z}^2\chi_\text{ee}^{{xx}^2}}
        \end{align*}
        \vspace{0.5em} 
    \end{minipage}
} \\ \cline{2-4}
\multicolumn{1}{|l|}{\raisebox{0.5em}[0pt]{\rotatebox{90}{Phase invariance}}} & 
\makebox[2cm][c]{
    \begin{minipage}{2cm}
        \vspace{0.5em} 
        \begin{align*}
            \chi_\text{ee}^{zz} & = \chi_\text{ee}^{xx} \\
            \chi_\text{mm}^{yy} & = -\chi_\text{ee}^{xx}
        \end{align*}
        \vspace{0.5em} 
    \end{minipage}
} & 
\makebox[2cm][c]{
    \begin{minipage}{2cm}
        \vspace{0.5em} 
        \begin{align*}
            T_\text{co}(\theta) & \in \mathbb{R} \\
            R_\text{co}(\theta) & \in \mathbb{I}
        \end{align*}
        \vspace{0.5em} 
    \end{minipage}
} &
\makebox[3.5cm][c]{
    \begin{minipage}{2cm}
        \vspace{0.5em} 
        \begin{align*}
            T_\text{co}(\theta) & = \frac{4 - k_\text{z}^2\chi_\text{ee}^{{zz}^2}}{4 + k_\text{z}^2\chi_\text{ee}^{zz2}} \\
            R_\text{co}(\theta) & = -\frac{4jk_\text{z}\chi_\text{ee}^{zz}}{4 + k_\text{z}^2\chi_\text{ee}^{{zz}^2}}
        \end{align*}
        \vspace{0.5em} 
    \end{minipage}
} \\ \hline
\multicolumn{1}{|l|}{\multirow{2}{*}{}} & 
\makebox[2.4cm][c]{
    \begin{minipage}{2cm}
        \vspace{0.5em} 
        \begin{align*}
            \chi_\text{ee}^{zz} & = 0 \\
            \chi_\text{mm}^{yy}\chi_\text{ee}^{xx} & = -\frac{4}{k^2}
        \end{align*}
        \vspace{0.5em} 
    \end{minipage}
} & 
\makebox[2cm][c]{
    \begin{minipage}{2cm}
        \vspace{0.5em} 
        \begin{align*}
            T_\text{co}(\theta) & = 0 \\
            |R_\text{co}(\theta)| & = 1
        \end{align*}
        \vspace{0.5em} 
    \end{minipage}
} & 
\makebox[3.5cm][c]{
    \begin{minipage}{2cm}
        \vspace{0.5em} 
        \begin{align*}
            T_\text{co}(\theta) & = 0 \\
            R_\text{co}(\theta) & = \frac{2 - jk_\text{z}\chi_\text{ee}^{xx}}{2 + jk_\text{z}\chi_\text{ee}^{xx}}
        \end{align*}
        \vspace{0.5em} 
    \end{minipage}
} \\ \cline{2-4} 
\multicolumn{1}{|l|}{\raisebox{-1em}[0pt]{\rotatebox{90}{Amplitude invariance}}} & 
\makebox[2cm][c]{
    \begin{minipage}{2cm}
        \vspace{0.5em} 
        \begin{align*}
            \chi_\text{ee}^{zz} & = -\chi_\text{ee}^{xx} \\
            \chi_\text{mm}^{yy} & = \chi_\text{ee}^{xx}
        \end{align*}
        \vspace{0.5em} 
    \end{minipage}
} & 
\makebox[2cm][c]{
    \begin{minipage}{2cm}
        \vspace{0.5em} 
        \begin{align*}
            |T_\text{co}(\theta)| & = 1 \\
            R_\text{co}(\theta) & = 0
        \end{align*}
        \vspace{0.5em} 
    \end{minipage}
} & 
\makebox[3.5cm][c]{
    \begin{minipage}{2cm}
        \vspace{0.5em} 
        \begin{align*}
            T_\text{co}(\theta) & = \frac{2 - jk_\text{z}\chi_\text{ee}^{xx}}{2 + jk_\text{z}\chi_\text{ee}^{xx}} \\
            R_\text{co}(\theta) & = 0
        \end{align*}
        \vspace{0.5em} 
    \end{minipage}
}
\label{table_summary_diagonal_anis}
\\ \hline
\end{tabular}
\end{table}
\subsubsection{\label{subsec:amplitude_inv} Angle-Invariant Amplitude}

To investigate the case of angle-invariant scattering amplitude, we follow a procedure similar as the one described for the case of angle-invariant phase discussed in the previous section. To obtain either a full transmission ($|T_\text{co}(\theta)| = 1$) or a full reflection ($|R_\text{co}(\theta)| = 1$), then either of them should have its real and imaginary parts equal to zero. For instance, to achieve full reflection, the real and imaginary parts of the transmission $T_\text{co}$ must be zero and thus~\eqqref{eq_condphaseinvT} must be satisfied. For this case, the solution is given by
\begin{equation}
    \chi_{\text{mm}}^{{yy}}(\theta) = -\frac{4 + k_x^2 \chi_{\text{ee}}^{{xx}}\chi_{\text{ee}}^{{zz}}}{\chi_{\text{ee}}^{{xx}}k^2}.
\end{equation}
This solution is again dependent on the wave vector component $k_x$, which is eliminated by setting $\chi_\text{ee}^{zz} = 0$. This leads to
\begin{subequations}
\begin{align}
    \chi_\text{ee}^{zz} &= 0, \\
    \chi_\text{mm}^{yy} \Big|_{\chi_{\text{ee}}^{zz} = 0}&= -\frac{4}{\chi_\text{ee}^{xx} k^2}.
    \end{align}    
\label{angle_indep_cond3}
\end{subequations}
This response represents an angle-invariant transmission amplitude, unaffected by the angle of incidence. Although the corresponding phase varies, as may be deduced from the scattering expressions provided in the third row of\tabref{table_summary_diagonal_anis}. Note that the conditions in~\eqqref{angle_indep_cond3} are compatible with those already discussed in~\cite{holloway2009a}.
The validity of this condition is demonstrated using the simulation in\figref{fig:spectrum_and_susceptibility}{(b-c)}, where the transmission is zero at the positions of the first and third dashed lines, corresponding to 11.5~GHz and 16.35~GHz. If the transmission is zero at these frequencies, then it necessarily implies that there is full reflection amplitude that is independent from the angle of incidence since the metasurface is made of lossless material. Additionally, referring to the insets of\figref{fig:spectrum_and_susceptibility}{(d)}, we see that the retrieved susceptibilities are in very good agreement with the conditions in~\eqqref{angle_indep_cond3}.
\newline 

We now consider the opposite effect, i.e., a full angle-invariant transmission in amplitude. In this case,~\eqqref{eq_condphaseinvR} must be satisfied, which leads to the condition
\begin{equation}
\label{eq_condAngInvTAmp}
    \chi_{\text{mm}}^{{yy}}(\theta) = \frac{k_z^2 \chi_{\text{ee}}^{{xx}} - k_x^2 \chi_{\text{ee}}^{{zz}}}{k^2}.
\end{equation}
The angular dependence of the solution is once more evident due to the terms $k_x^2$ and $k_z^2$. To eliminate this angular dependence and achieve a fully angle-independent solution, we set $\chi_\text{ee}^{xx} = -\chi_\text{ee}^{zz}$, which reduces~\eqqref{eq_condAngInvTAmp} to
\begin{subequations}
\begin{align}
\chi_\text{ee}^{xx} &= -\chi_\text{ee}^{zz}, \\
\chi_\text{mm}^{yy} \Big|_{\chi_{\text{ee}}^{zz} = -\chi_\text{ee}^{zz}}&= \chi_\text{ee}^{zz}.
\end{align}
\label{eq_cond_AngInvTAmp2}
\end{subequations}
These conditions result in a transmission that is fully independent of the angle of incidence for the amplitude, while the phase retains its angular dependence, as can be deduced from the expression of the scattering parameters provided in the fourth row of\tabref{table_summary_diagonal_anis}.

\begin{figure}[h]
    \centering
    \includegraphics[width=\linewidth]{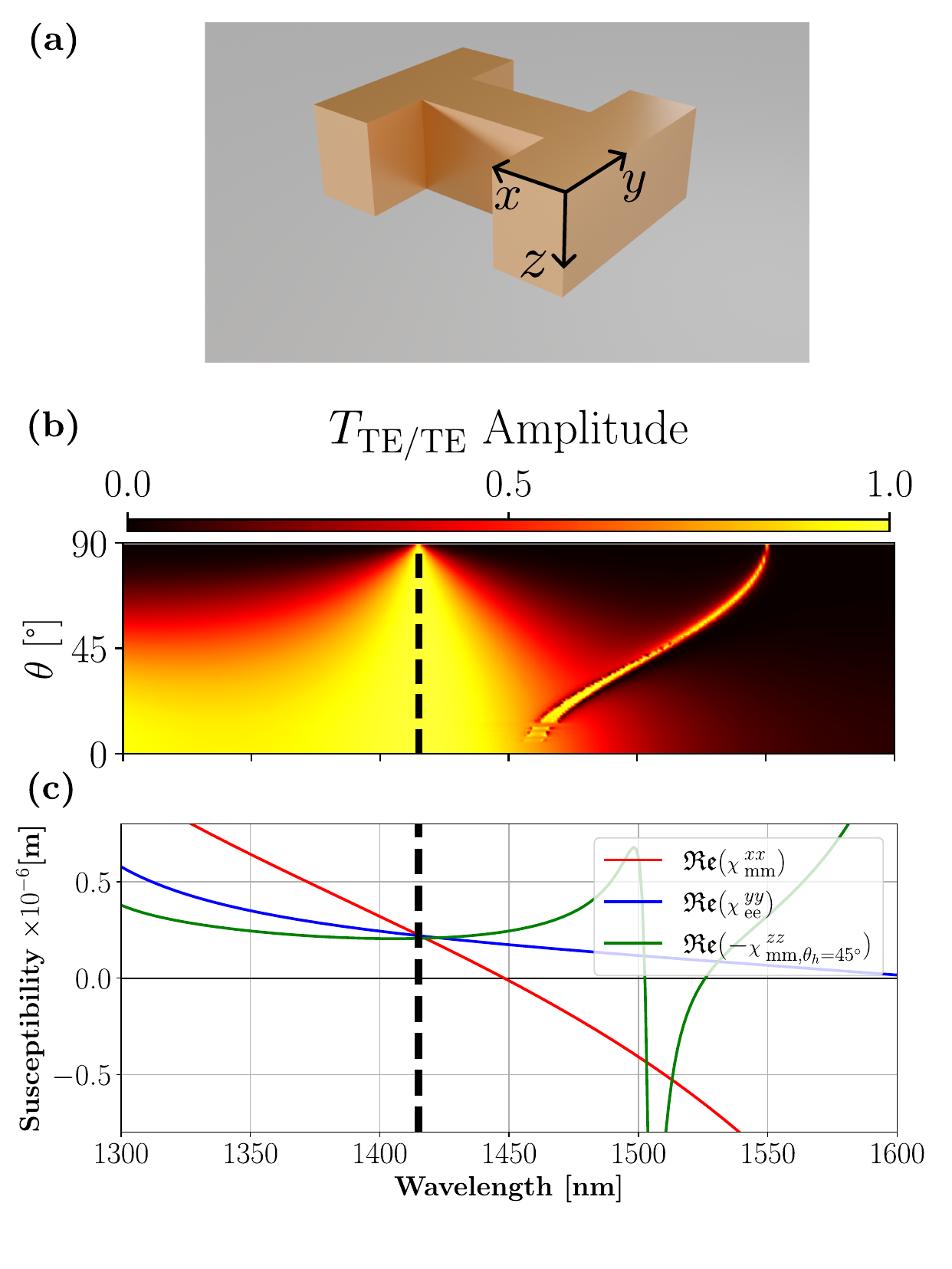}
    \caption{\textbf{Dog-bone structure simulation in the optical regime.} \textbf{(a)} Geometry of the unit cell. \textbf{(b)} Simulation of the TE co-polarized transmission coefficient for different angles and frequencies ranging from $0^\circ$ to $90^\circ$ and $1300$ nm to $1600$ nm, respectively. \textbf{(c)} Retrieved susceptibilities using the simulated fields and the GSTC model. }
    \label{fig:spectrum_and_susceptibility_singleH}
\end{figure}

To demonstrate that a non-trivial angle-independent transmission amplitude is possible, we consider an optical metasurface formed by a periodic arrangement of the \mbox{H-shaped} particle depicted in\figref{fig:spectrum_and_susceptibility_singleH}{(a)}. The physical parameters of this metasurface are provided in \textbf{Appendix E} of the SI. The resulting full-wave simulated TE transmission amplitude is shown in\figref{fig:spectrum_and_susceptibility_singleH}{(b)} from which we can clearly see that the amplitude is angle-invariant at a wavelength of 1415~nm. Again, to verify that this response matches the condition in~\eqqref{eq_cond_AngInvTAmp2}, we calculate the metasurface effective susceptibilities following the procedure described in \textbf{Appendix F1} of the SI. Note that since we are considering TE polarization instead of TM, the relevant susceptibilities are $\chid{mm}{xx}$, $\chid{mm}{zz}$ and $\chid{ee}{yy}$. The resulting susceptibilities are plotted in\figref{fig:spectrum_and_susceptibility_singleH}{(c)} from which we see that the conditions in~\eqqref{eq_cond_AngInvTAmp2} are indeed satisfied at $\lambda=1415$~nm.
\subsection{Off-Diagonal Anisotropic Metasurfaces}
We now investigate the scattering responses of purely off-diagonal anisotropic metasurfaces. To simplify our study, we consider cases where the diagonal elements are assumed to be null, such that 
\begin{subequations}
\label{off-diag-susceptbilities}
\begin{alignat}{2}
    \chi_\text{ee,mm}^{ii} &= 0, &\quad i &\in \{x, y, z\}, \\
    \chi_\text{ee,mm}^{ij} &\neq 0, &\quad i, j &\in \{x, y, z\}, \quad i \neq j.
\end{alignat}
\end{subequations}
We recall again that $\te{\chi}\@{me} = \te{\chi}\@{em} = 0$ because we are still considering anisotropic metasurfaces. The transmission and reflection coefficients for most susceptibility combinations of \eqqref{off-diag-susceptbilities} are angle-dependent, which is not desirable. The following subsections explore conditions for angle-independent scattering in two off-diagonal anisotropic metasurfaces.
\subsubsection{Angle-Independent Transmission Leading to Spatial Differentiation}
We now discuss angular invariance in both the $xz$- and $yz$-planes and for both the amplitude and phase of the scattering coefficients. For simplicity, let us first consider that the only nonzero metasurface susceptibilities are $\chid{ee}{yz}$ and $\chid{mm}{yz}$. The co- and cross-polarized transmission and reflection coefficients have been derived for both polar and azimuthal angles $\theta$ and $\phi$, respectively, and are shown in~\textbf{(C15)-(C18)} of the SI. 

\begin{figure*}[ht]
    \centering
    \includegraphics[width=\textwidth]{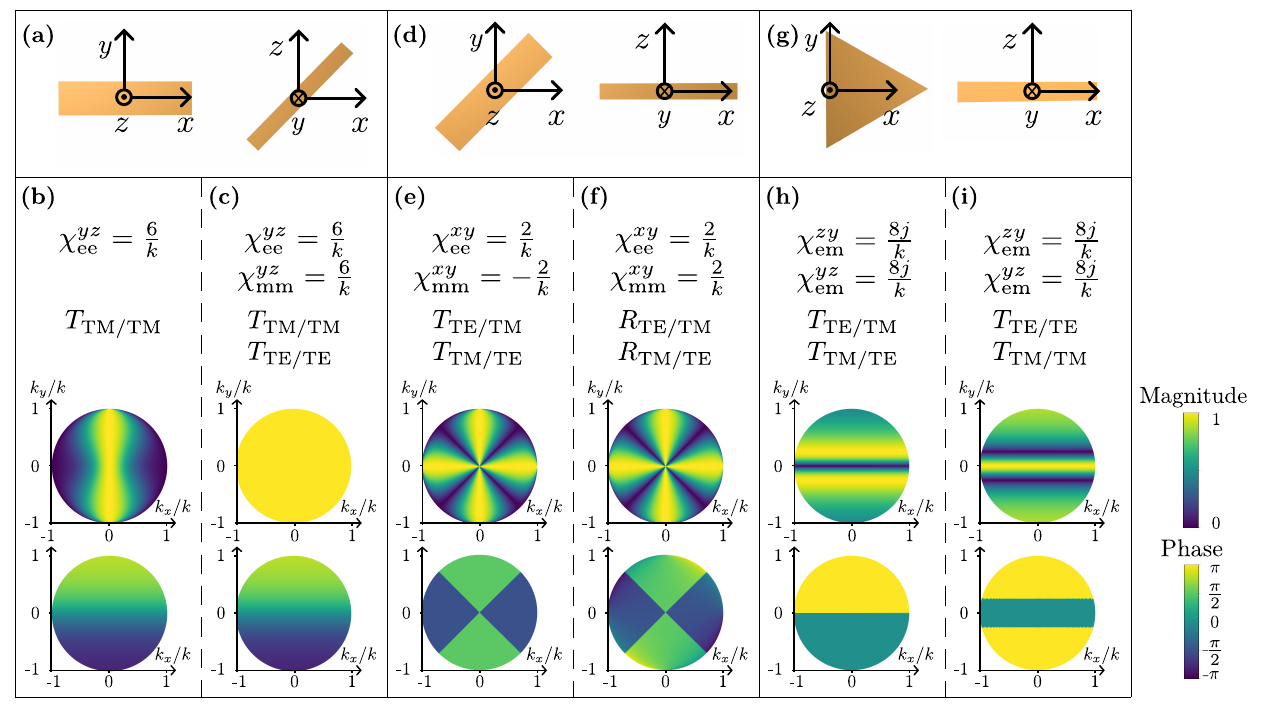}
    \caption{\textbf{Fourier-plane plots of the scattering parameters corresponding to different unit-cell and angle-invariant conditions.} Top row: geometry of the unit cells that may allow implementing, from the perspective of spatial symmetries, the considered angle-invariant conditions. Second row: considered susceptibility conditions. Bottom row: Fourier-plan magnitude and phase plots corresponding to the considered conditions and the specified scattering parameters. \textbf{(a)}~A structure presenting both a $180^\circ$-rotation symmetry ($C_{2,y}$) and a reflection symmetry ($\sigma_y$) along the $y$-axis would exhibit nonzero $\chid{ee}{yz}$ and $\chid{mm}{yz}$. If, by optimizing the structure, we achieve $\chid{mm}{yz}= 0$ and $\chid{ee}{yz} \neq 0$, then an angle-invariant co-transmission amplitude and an angle-varying asymmetric phase shift is obtained along the $yz$-plane~\textbf{(b)}. However, if we are able to achieve $\chid{mm}{yz}=\chid{ee}{yz} \neq 0$, then the co-transmission amplitude becomes angle-invariant within the full visible light cone~\textbf{(c)}. \textbf{(d)}~A structure presenting both a $180^\circ$-rotation symmetry ($C_{2,z}$) and a reflection symmetry ($\sigma_z$) along the $z$-axis would exhibit nonzero $\chid{ee}{xy}$ and $\chid{mm}{xy}$. If either of the conditions $\chid{ee}{xy} = \mp \chid{mm}{xy} \neq 0$ are met, then a total angle-invariant cross-polarized transmission~\textbf{(e)}, respectively reflection~\textbf{(f)},  may be obtained within both the $xz$- and the $yz$-planes. \textbf{(g)}~A structure being reflection symmetric along both the $y$- and the $z$-axes ($\sigma_y$ and $\sigma_z$) would exhibit nonzero $\chid{em}{yz}$ and $\chid{em}{zy}$. If the condition $\chid{em}{yz}=\chid{em}{zy}=8j/k$ is satisfied, then co- and cross-polarized $k_x$-independent full transmissions with binary phase shifts are obtained for two specific values of $k_y$~\textbf{(h,i)}. Note that in the examples~\textbf{(a)} and~\textbf{(d)}, the values attributed to the susceptibilities were chosen arbitrarily but, nonetheless, such that the metasurfaces remain passive and lossless according to~\eqqref{eq_recipLP}.}
    \label{fig_fourier_planes}
\end{figure*}

Let us start our analysis with the special case where a metasurface with $\chid{mm}{yz}=0$ is illuminated by a TE-polarized plane wave. In this case, the co-polarized transmission is given by
\begin{align}
\label{eq_TcoXeeyz}
   T_{\text{co}}(\theta, \phi)= \frac{k_z^2(4+k_y^2\chid{ee}{{yz}^2})}{4k_z^2+(k^2k_x^2+k_y^2k_z^2)\chid{ee}{{yz}^2}}.
\end{align}
This expression is particularly interesting since it reduces to $|T_{\text{co}}(\theta, \phi)|=1$ when $k_x=0$, thus providing angle-invariant amplitude within the $xz$-plane. As an illustration, the Fourier plane response of the transmission coefficient in~\eqqref{eq_TcoXeeyz} is plotted in\figref{fig_fourier_planes}{(b)}. This demonstrates that such a metasurface indeed exhibits an angle-independent transmission amplitude within the $yz$-plane. Interestingly, this metasurface also induces an angle-asymmetric phase shift relative to $k_y$, which, as we will see shortly, may be linked to spatial differentiation. 
\newline

Another interesting case is achieved with the specific condition
\begin{align}
\label{eq_kerker1}
    \chid{ee}{yz} = \chid{mm}{yz},
\end{align}
which is reminiscent of the Kerker condition in~\eqqref{eq_kerker}. If this condition is satisfied, then the cross-polarized transmission coefficient is nullified as well as the reflection coefficient, leading to
\begin{subequations}
\label{off_diag_anis_1}
\begin{align}
    T\@{co}(k_x=0, k_y) &= \frac{2j - k_y \chid{ee}{yz}}{2j + k_y \chid{ee}{yz}}, \\  R\@{co}(k_x=0, k_y)&=R\@{cr}(k_x=0, k_y)=0.
\end{align}
\end{subequations}
Despite the presence of the wave vector component $k_y$ in the expression of the transmission coefficient in~\eqqref{off_diag_anis_1}, it turns that the latter exhibits a full angle-invariant amplitude, i.e., $|T\@{co}(\theta, \phi)| = 1$. Moreover, the corresponding transmission phase only changes in the $yz$-plane. To illustrate the angular response of~\eqqref{off_diag_anis_1}, we plot the Fourier plane amplitude and phase of the transmission coefficient in\figref{fig_fourier_planes}{(c)}. Note that if the considered susceptibilities are $\chid{ee}{xz}$ and $\chid{mm}{xz}$, applying a similar condition $\chid{ee}{xz} = \chid{mm}{xz}$ allows us to rotate the Fourier plane by $90^\circ$ (the phase variation would occur along the $k_x$ direction). In this case, the response is independent of the polarization (TE or TM), as both polarizations share the same Fourier plane behavior. 

From the results shown in\figref{fig_fourier_planes}{(a)}, it is apparent that the transmission has almost a linear phase change over $k_y$. This is verified by taking a first-order Taylor expansion of the transmission coefficient in~\eqqref{off_diag_anis_1} with respect to $k_y$, which yields
\begin{align}
    T\@{co}(0,k_y) \approx T\@{co}(0,0) + k_y \frac{\partial T\@{co}(0,k_y)}{\partial k_y}\Big|_{k_y=0},
\end{align}
where $T\@{co}(0,0) = 1$ and $\frac{\partial T\@{co}(0,k_y)}{\partial k_y}\Big|_{k_y=0} = j\chid{ee}{yz}$. For small values of $k_y\chid{ee}{yz}$, the argument of this transmission coefficient may therefore be approximated as a linear function of $k_y$ as
\begin{align}
    \angle T\@{co}(0,k_y) = \tan^{-1}\left ( k_y\chid{ee}{yz} \right ) \approx k_y\chid{ee}{yz}.
\end{align}
This could be seen as a spatial derivative of the transmission coefficient that is imaged as a linear phase shift in the Fourier domain\rcite{doskolovich2014spatial,momeni2019}. Therefore, if a metasurface with is designed to exhibit only $\chid{ee}{yz}$ (the case with $\chid{mm}{yz}$ is too difficult to practically implement), then the spatial differentiation would only occur within the $yz$-plane, as shown in\figref{fig_fourier_planes}{(b)}. However, if the ``generalized'' Kerker condition in~\eqqref{eq_kerker1} is satisfied, then the spatial differentiation would occur within the full Fourier plane, as shown in\figref{fig_fourier_planes}{(c)}.
\subsubsection{Partial Gyrotropic Angular Invariance}
\label{sec_AIPR}
Let us now consider a metasurface with the only nonzero susceptibilities being $\chi\@{mm}^{xy}$ and $\chi\@{ee}^{xy}$. By solving~\eqqref{eq_GSTC} along with~\eqqref{eq_PM}, we once again obtain the transmission and reflection coefficients. The resulting expressions are lengthy and, for convenience, we only show that they are proportional to (refer to the SI for the complete expressions)
\begin{subequations}
\begin{align}
     R\@{co}(\theta) &\propto (\chid{mm}{xy} + \chid{ee}{xy})(\chid{mm}{xy} - \chid{ee}{xy}), \\ 
    R\@{cr}(\theta) &\propto (\chid{ee}{xy} + \chid{mm}{xy})(4 + k^2\chid{ee}{xy}\chid{mm}{xy}), \\
    T\@{co}(\theta) &\propto (4 + k^2\chid{ee}{xy}\chid{mm}{xy})(4 - k^2\chid{ee}{xy}\chid{mm}{xy}), \\
    T\@{cr}(\theta) &\propto (\chid{ee}{xy} - \chid{mm}{xy})(k^2\chid{ee}{xy}\chid{mm}{xy} - 4).
\end{align}
\label{off_diag_anis2}
\end{subequations}
Due to the presence of the tangential off-diagonal susceptibilities, $\chi\@{mm}^{xy}$ and $\chi\@{ee}^{xy}$, we now have both co- and cross-polarized responses. Note that these equations exhibit only partial angular variation that is limited to either the $xz$- or the $yz$-planes. 
\newline 

The first interesting condition that may be deduced from \eqqref{off_diag_anis2} is given by
\begin{align}
    \chid{ee}{xy} = -\chid{mm}{xy}.
\end{align}
If this condition is satisfied, then both co- and cross-polarized reflection coefficients are zero irrespectively of the incidence angle. This condition corresponds to a form of generalized counter-part of the Kerker condition in~\eqqref{eq_kerker} that applies for all angles instead of being restricted to normally incident waves. If we further impose that $\chid{ee}{xy} = -\chid{mm}{xy} = \mp 2/k$, then it becomes possible to not only nullify the reflection but also the co-polarized transmission. It follows that, within the $xz$- and $yz$-planes (for $\phi = \{0,\pi/2\}$), the resulting scattering coefficients reduce to
\begin{equation}
\label{off_diag_anis3}
\begin{split}
    R\@{co}(\theta) = 0, &\qquad R\@{cr}(\theta) = 0, \\
    T\@{co}(\theta) = 0, &\qquad T\@{cr}(\theta) = \pm j.
\end{split}
\end{equation}
In~\eqqref{off_diag_anis3}, the only remaining term is a purely imaginary cross-polarized transmission leading to a phase shift of $\pm\pi/2$, and a corresponding full amplitude. We see here the existence of an interesting trade-off: canceling the co-polarized transmission by setting the susceptibilities equal to $\mp 2/k$ allows achieving a full-amplitude cross-polarized transmission at the cost of a limited binary phase shift of $\pm\pi/2$. Nevertheless, the cross-polarization transmission does not depend on the angle of incidence, making it a fully angle-independent polarization converter, at least within the $xz$- and $yz$-planes. An example of such a transmission is illustrated in\figref{fig_fourier_planes}{(e)}.
\newline

Alternatively, we may consider the condition
\begin{align}
    \chid{ee}{xy} = \chid{mm}{xy},
\end{align}
which, despite being reminiscent of the Kerker condition in~\eqqref{eq_kerker}, leads to a quite different outcome. Indeed, this condition allows canceling both the cross-polarized transmission and the co-polarized reflection, i.e., \mbox{$T\@{cr}=R\@{co}=0$}. By further imposing that \mbox{$\chid{ee}{xy} = \chid{mm}{xy} = \pm 2/k$}, we may, this time, nullify the co-polarized transmission to be left with only
\begin{equation}
\label{off_diag_anis4}
    \begin{split}
        R\@{co}(\theta) = 0, & \qquad R\@{cr}(\theta) = \pm j, \\
        T\@{co}(\theta) = 0, & \qquad T\@{cr}(\theta) = 0.
    \end{split}  
\end{equation}
We can clearly notice that the cross-polarized reflection is the only term remaining and that it is angle invariant. Once again, we find a trade-off similar to the one discussed above but that, this time, applies to the phase of the cross-polarized reflection. This case is illustrated in\figref{fig_fourier_planes}{(f)}.

Overall, both cases shown in\figref{fig_fourier_planes}{(d)} demonstrate that angle-invariant polarization conversion with full amplitude is possible but is limited to the $xz$- and $yz$-planes. Note that the type of anisotropy described in this section corresponds to birefringence. Consequently, the polarization conversion operation illustrated in\figref{fig_fourier_planes}{(d)} does not stem from chiral responses but rather from the birefringent nature of the metasurfaces that act as angle-invariant half-wave plates operating either in transmission or reflection.

\subsection{\label{subsec:total_inv}Bianistropic Metasurfaces}

We now investigate angle-invariant scattering in the general case of bianisotropic metasurfaces, for which none of the susceptibility tensors in~\eqqref{eq_PM} are zero. While there may be several combinations of susceptibilities that lead to partial amplitude and/or phase angular invariance, we shall concentrate our attention on only three particularly interesting cases.

\subsubsection{\label{nongyrotropic}Complete Nongyrotropic Angular Invariance}

We have seen in Sec.~\ref{sec_AIPR} that full-amplitude angular invariance, within either the $xz$- or $yz$-plane, is possible but only for cross-polarized transmission or reflection. We shall now consider a situation where full-amplitude angular invariance may be achieved in a nongyrotropic fashion (without polarization rotation). This case was introduced in~\cite{lavigne2019,achouri2021}, and is briefly revisited here for completeness.

Let us consider a reciprocal metasurface for which the only nonzero susceptibilities are $\chid{ee}{xx}$, $\chid{mm}{yy}$ and $\chid{em}{xy}$. The resulting scattering parameters for TM-polarized waves propagating in the $xz$-plane are given by
\small
\begin{subequations}
\label{eq_RTbiani}
\begin{align}
T_{\text{co}}(\theta) &= \frac{-j k_{z} \left[\frac{k^2}{2} (\chi_{\text{em}}^{xy^2} + \chi_{\text{ee}}^{{xx}} \chi_{\text{mm}}^{{yy}})+2\right]}{k_{z}^2 \chi_{\text{ee}}^{{xx}} + k^2 \chi_{\text{mm}}^{{yy}} + j k_{z} \left[ \frac{k^2}{2} (\chi_{\text{em}}^{xy^2} + \chi_{\text{ee}}^{{xx}} \chi_{\text{mm}}^{{yy}})-2\right]},
\label{eq:BB2}\\
R_{\text{co}}(\theta) &= \frac{- k_{z}^2 \chi_{\text{ee}}^{{xx}} \mp 2 k k_{z} \chi_{\text{em}}^{xy} + k^2 \chi_{\text{mm}}^{{yy}}}{k_{z}^2 \chi_{\text{ee}}^{{xx}} + k^2 \chi_{\text{mm}}^{{yy}} + j k_{z} \left[ \frac{k^2}{2} (\chi_{\text{em}}^{xy^2} + \chi_{\text{ee}}^{{xx}} \chi_{\text{mm}}^{{yy}})-2\right]} \label{eq:BB1},
\end{align}
\end{subequations}
\normalsize
where the $\mp$ sign in~\eqqref{eq:BB1} indicates that this reflection coefficient is different when the incident wave exciting the metasurface propagates in the $+z$-direction (top sign) or in the $-z$-direction (bottom sign). This difference in sign in the reflection coefficient is a consequence of the fact that a metasurface exhibiting a nonzero $\chid{em}{xy}$ susceptibility must necessarily be asymmetric $z$-direction~\cite{achouri2021,achouri2023a}. It can be clearly seen that the~\eqqref{eq:BB2} depends on the angular variable $k_z$, which itself depends upon $k_x$. In order to remove this angular dependency, one can set the following conditions
\begin{subequations}
\label{eq_XeexxXmmyyCond}
\begin{align}
    \chi\@{ee}^{xx} = \chi\@{mm}^{yy} = 0,\label{eq_XeexxXmmyyCond1} \\
    \chi\@{em}^{xy} \neq 0,
\end{align}
\end{subequations}
which may be practically achieved within a narrow frequency band by exploiting multiple Lorentzian resonances such that both electric and magnetic susceptibilities approach zero, as similarly done for the case in\figref{fig:spectrum_and_susceptibility}. If the conditions in~\eqqref{eq_XeexxXmmyyCond} are satisfied, then the transmission and reflection coefficients in~\eqqref{eq_RTbiani} reduce to
\begin{subequations}
\label{eq_chiemxyTR}
\begin{align}
    T\@{co}(\theta) &= \frac{4 + k^2 \chi\@{em}^{{xy}^2}}{4 - k^2 \chi\@{em}^{{xy}^2}}, \label{eq:chiemxyT} \\
    R\@{co}(\theta) &= \frac{4jk \chi\@{em}^{xy}}{4 - k^2 \chi\@{em}^{{xy}^2}}.\label{eq:chiemxyR}
\end{align}
\end{subequations}
These relations clearly show that such a metasurface exhibits angle-invariant co-polarized transmission and reflection. An important consequence that may be deduced from this example is that a metasurface satisfying~\eqqref{eq_XeexxXmmyyCond} would correspond to a purely bianisotropic metasurface and, since the susceptibility $\chi\@{em}^{xy}$ is associated to nonlocal (spatially dispersive) responses~\cite{achouri2023a}, such a metasurface would thus be purely nonlocal. Importantly, this also demonstrates that nonlocality does \emph{not} only increase the angular dispersion of a metasurface scattering response but may also be used as a mean to completely remove all of its angular dependence, as demonstrated by the angle-independent scattering parameters in~\eqqref{eq_chiemxyTR}.

Upon inspection of~\eqqref{eq_chiemxyTR}, it clearly appears that the reflection amplitude may be maximized by selecting \mbox{$\chid{em}{xy} = \pm 2j/k$}. If this additional condition is satisfied, then the transmission coefficient becomes $T\@{co}(\theta) = 0$ and the reflection coefficient reduces to $R\@{co}(\theta) = \pm 1$. Note that this condition forces the susceptibility $\chid{em}{xy}$ to be purely imaginary, which is consistent with the prescription of passivity and losslessness given in~\eqqref{eq_recipLP}. 

\begin{figure}[H]
    \centering
    \includegraphics[width=\linewidth]{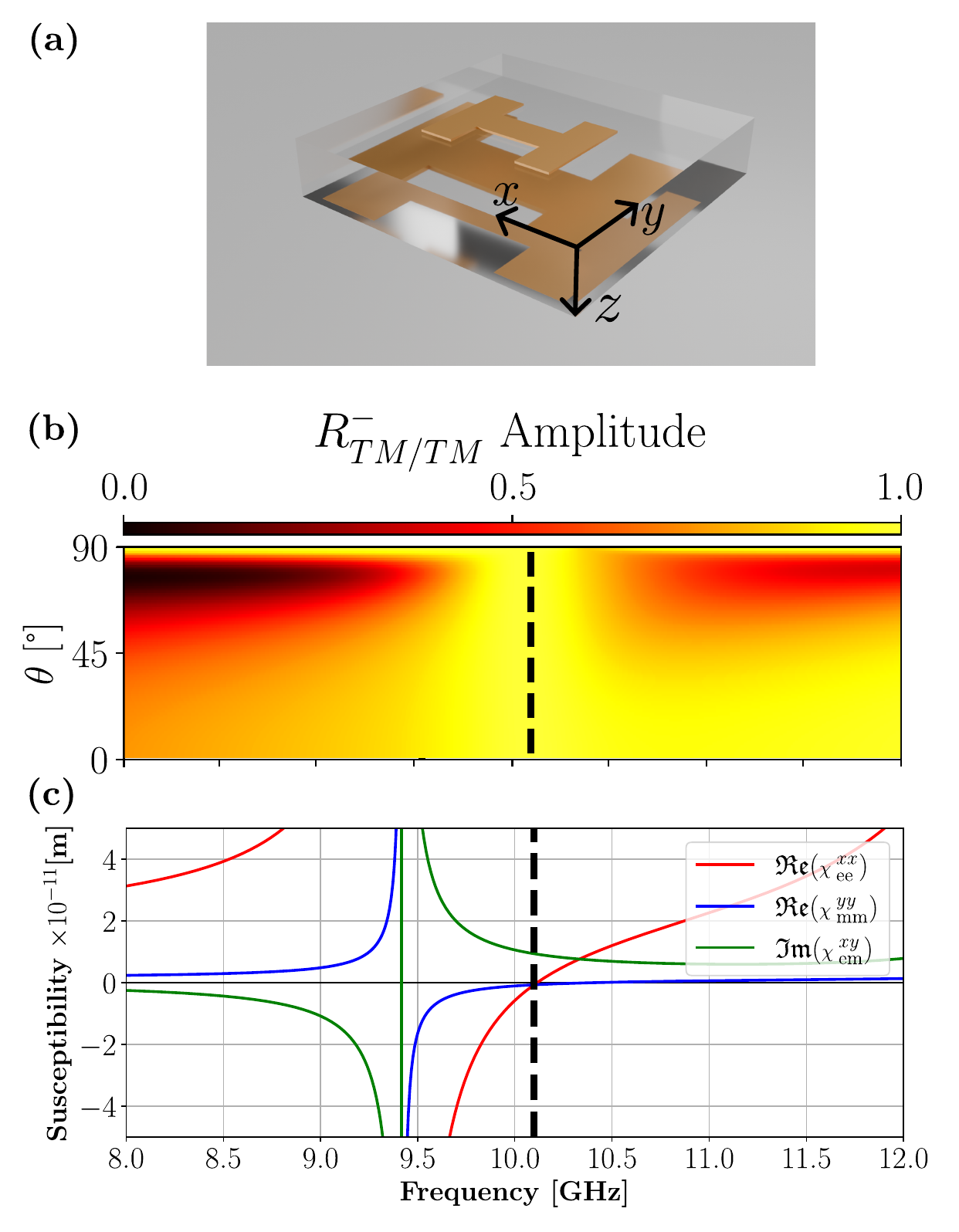}
    \caption{\textbf{Dog-bone structure simulation in the microwave regime.} \textbf{(a)} Geometry of the unit cell. Notice that the dog-bone at the top layer has different dimensions to the one at the bottom layer. \textbf{(b)} Simulation of the TM co-polarized backward reflection coefficient for different angles and frequencies ranging from $0^\circ$ to $90^\circ$ and $8$ GHz to $12$ GHz, respectively. \textbf{(c)} Retrieved susceptibilities using the simulated fields and the GSTC model.}
    \label{fig_spectrum_and_susceptibility_doubleH}
\end{figure}

An example of a structure that satisfies the conditions in~\eqqref{eq_XeexxXmmyyCond} along with $\chid{em}{xy} = 2j/k$ is provided in\figref{fig_spectrum_and_susceptibility_doubleH}{}. As explained above, the existence of the susceptibility $\chid{em}{xy}$ is obtained by breaking the reflection symmetry of a structure in the $z$-direction~\cite{achouri2021,achouri2023a}. For this reason, an asymmetric double dog-bone structure is used, as illustrated\figref{fig_spectrum_and_susceptibility_doubleH}{(a)}. Its TM co-polarized backward (incident wave propagation in the $-z$-direction) reflection amplitude is plotted in\figref{fig_spectrum_and_susceptibility_doubleH}{(b)}, where we can see that it reaches an angle-invariant $|R_\text{co}^-(\theta)|=1$ at 10.1~GHz. The retrieved metasurface susceptibilities plotted in\figref{fig_spectrum_and_susceptibility_doubleH}{(c)} confirm that the conditions in~\eqqref{eq_XeexxXmmyyCond} are indeed achieved along with $\chid{em}{xy} = 2j/k$ near 10.1~GHz.

Our discussion has so far been restricted to the susceptibility $\chid{em}{xy}$. However, in a general case, a metasurface that exhibits a nonzero $\chid{em}{xy}$ susceptibility would also typically exhibit a nonzero $\chid{em}{yx}$. The existence of $\chid{em}{yx}$ would not change the metasurface scattering response in the $xz$-plane (assuming TM polarization). So, even if $\chid{em}{yx}\neq 0$, Eqs.~\eqref{eq_RTbiani} and~\eqref{eq_chiemxyTR} would remain unchanged. Nevertheless, the existence of $\chid{em}{yx}$ would affect the scattering response of the metasurface for an arbitrary direction of wave propagation. In such a case, we highlight that the condition
\begin{equation}
   \label{eq_condbianiFP}
    \chid{em}{xy} = -\chid{em}{yx},
\end{equation}
leads to total angle invariance in the entire Fourier plane. If this condition is satisfied, in addition to~\eqqref{eq_XeexxXmmyyCond}, then the metasurface scattering parameters would be identical to those in~\eqqref{eq_chiemxyTR} but these expressions would apply for an arbitrary set $\{\theta,\phi\}$ of propagation angles. Note that the condition of~\eqqref{eq_condbianiFP} is trivial to satisfy since the only requirements are that the metasurface is asymmetric in the $z$-direction while being reflection symmetric in the $x$- and $y$-directions ($\sigma_x$ and $\sigma_y$) and presents a $90^\circ$-rotation symmetry along the $z$-axis ($C_{4,z}$)~\cite{achouri2023a}. The real challenge lies in achieving the condition of~\eqqref{eq_XeexxXmmyyCond1}.

\subsubsection{Angle-Invariant Circular Polarization Conversion}

We now expand the theory related to the susceptibilities $\chid{em}{xy}$ and $\chid{em}{yx}$ to the angle-independent scattering of circularly polarized (CP) waves. Let us start by expressing the transmitted and reflected fields in the circular polarization basis as a function of the linear polarized ones and the corresponding transmission and reflection scattering matrices as (refer to \textbf{Appendix D2} in the SI for the details)
\begin{subequations}
\label{change_basis}
\begin{align}
    \ve{E}\@{CP}^\text{t} &= \underbrace{\te{\Lambda}\@{2} \cdot\; \te{T}\@{LP} \cdot\; \te{\Lambda}\@{1}^{-1}}_{\te{T}\@{CP}} \cdot \ve{E}\@{CP}^\text{i}, \\
    \ve{E}\@{CP}^\text{r} &= \underbrace{\te{\Lambda}\@{3} \cdot\; \te{R}\@{LP} \cdot\; \te{\Lambda}\@{1}^{-1}}_{\te{R}\@{CP}} \cdot \ve{E}\@{CP}^\text{i},
\end{align}
\end{subequations}
where we define $ \ve{E}\@{CP}^\text{t} = \left [E\@{RCP}^\text{t} \quad E\@{LCP}^\text{t}\right ]^T $ the transmitted and \mbox{$ \ve{E}_\text{CP}^\text{r} = \left [E\@{RCP}^\text{r} \quad E\@{LCP}^\text{r}\right ]^T $} the reflected electric fields in the CP basis; and $ \ve{E}\@{CP}^\text{i} = \left [E\@{RCP}^\text{i} \quad E\@{LCP}^\text{i}\right ]^T $ is the incident electric field in the same basis, with  RCP and LCP representing the right and left hand circular polarizations (RCP and LCP, respectively). The matrices $\te{T}\@{CP} $ and $\te{R}\@{CP} $ are the transmission and reflection matrices in the CP basis.
The matrix $ \te{\Lambda}\@{1} $ ($ \te{\Lambda}\@{2} $) is a transformation matrix that converts the incident (transmitted) electric field from the linear polarization (LP) basis to the CP basis, and its inverse $ \te{\Lambda}_1^{-1} $ performs the inverse operation. Similarly, the matrix $ \te{\Lambda}\@{3} $ converts reflected fields from the LP to the CP basis. These matrices, based on the Jones vector~\cite{Azzam_Bashara_1977}, are connected together via
\begin{equation}
    \te{\Lambda}\@{1} = \te{\Lambda}\@{2} = \te{\Lambda}\@{3}^* = \frac{1}{\sqrt{2}}
    \begin{bmatrix}
    1 & -j \\
    1 & j
    \end{bmatrix}.
\end{equation}
The matrices $\te{T}_\text{LP} $ and $ \te{R}_\text{LP} $ represent the transmission and reflection coefficients in the LP basis, and are respectively given by
\begin{subequations}
\label{eq_TLPLCP}
\begin{align}
    \te{T}\@{LP} &=
    \begin{bmatrix}
    T\@{TE/TE} & T\@{TE/TM} \\
    T\@{TM/TE} & T\@{TM/TM}
    \end{bmatrix},\\
    \te{R}\@{LP} &=
    \begin{bmatrix}
    R\@{TE/TE} & R\@{TE/TM} \\
    R\@{TM/TE} & R\@{TM/TM}
    \end{bmatrix}.
\end{align}
\end{subequations}
Here, the subscripts $ \rm TE $ and $ \rm TM $ denote Transverse Electric and Transverse Magnetic polarizations, while the terms $ T\@{TE/TE} $, $ T\@{TE/TM} $, $ R\@{TM/TM} $, and $ R\@{TE/TM} $, represent the specific transmission and reflection coefficients for each polarization state. Similarly, in the CP basis, we have
\begin{subequations}
\begin{align}
    \te{T}\@{CP} &=
    \begin{bmatrix}
    T\@{RCP/RCP} & T\@{RCP/LCP} \\
    T\@{LCP/RCP} & T\@{LCP/LCP}
    \end{bmatrix},\\
    \te{R}\@{CP} &=
    \begin{bmatrix}
    R\@{RCP/RCP} & R\@{RCP/LCP} \\
    R\@{LCP/RCP} & R\@{LCP/LCP}
    \end{bmatrix}.
\end{align}
\label{eq_TCPRCP}
\end{subequations}
We are now interested in realizing an angle-invariant polarization converter within the CP basis (e.g., from RCP to LCP). To do so, we set the diagonal terms of the matrices $\te{R}\@{CP}$ and $\te{T}\@{CP}$ to zero so as to remove the co-polarized scattering coefficients. Secondly, to obtain an angle-invariant response, the co- and cross-transmission, and the co- and cross-reflection coefficients in the LP basis should all be angle invariant. For this purpose, we choose the susceptibilities $\chi_\text{em}^{xy}$ and $\chi_\text{em}^{yx}$ that lead to angle-invariant responses only for the co-polarization coefficients in the LP basis and zero scattering for the cross-polarization ones, as explained in the SI and as shown in~\eqqref{eq_chiemxyTR}. Under these conditions, the scattering matrices in the CP basis reduce to
\begin{subequations}
\label{eq_TRCP}
\small
\begin{align}
\te{T}_{\text{CP}} &= 
\frac{1}{2}
\begin{bmatrix}
T_{\text{TM/TM}} + T_{\text{TE/TE}} &  
T_{\text{TM/TM}} - T_{\text{TE/TE}} \\
T_{\text{TM/TM}} - T_{\text{TE/TE}} & 
T_{\text{TM/TM}} + T_{\text{TE/TE}}
\end{bmatrix},\\
\te{R}_{\text{CP}} &= 
\frac{1}{2}
\begin{bmatrix}
R_{\text{TM/TM}} - R_{\text{TE/TE}} &  
R_{\text{TM/TM}} + R_{\text{TE/TE}} \\
R_{\text{TM/TM}} + R_{\text{TE/TE}} & 
R_{\text{TM/TM}} - R_{\text{TE/TE}}
\end{bmatrix}.
\end{align}   
\end{subequations}
Now, in order to retain only the off-diagonal elements of these CP scattering matrices, the following conditions must be satisfied
\begin{subequations}
\begin{align}
T\@{TE/TE} &= -T\@{TM/TM},\\
R\@{TE/TE} &= R\@{TM/TM}.
\end{align}
\label{cond_cp_anginv}
\end{subequations}
In which case,~\eqqref{eq_TRCP} become
\begin{subequations}
\label{tcprcp}
\begin{align}
\te{T}_{\text{CP}} &= 
-T_{\text{TM/TM}} \te{J}_2,\\
\te{R}_{\text{CP}} &= 
R_{\text{TM/TM}} \te{J}_2,
\end{align}
\end{subequations}
where $\te{J}_2$ is the exchange matrix\rcite{horn2013matrix}, while the co-polarized transmission and reflection coefficients are provided in Sec.~\textbf{D2} of the SI. Now, in order to satisfy the conditions in~\eqqref{cond_cp_anginv}, the only possibility is that 
\begin{align}
    \chi\@{em}^{xy}\chi\@{em}^{yx}  = \pm \frac{4}{k^\text{2}}.
    \label{sol_tcprcp}
\end{align}
If the condition of~\eqqref{sol_tcprcp} is satisfied, then the metasurface behaves as an angle-invariant circular polarization converter but only within the $xz$- and $yz$-planes. Moreover, such conversion occurs simultaneously in both transmission and reflection, which necessarily limits the efficiency in either channel. The condition in~\eqqref{sol_tcprcp} enables tuning both the phase and amplitude of reflection and transmission, which may be advantageous for specific applications.

\subsubsection{Binary-Phase Angle-Invariant Pseudochirality}

We now show that it is possible to design a metasurface with either full co- or cross-polarized transmission depending on the angle of incidence ($\theta,\phi$). We show that polarization conversion with unity amplitude can be achieved, independent of $k_x$ and for any chosen $k_y$.

To achieve such a response, the metasurface must exhibit nonzero $\chid{em}{zy}$ and $\chid{em}{yz}$. Note these susceptibilities are allowed to exist, from the perspective of spatial symmetries, when the scattering particles have a broken reflection symmetry in the $x$-direction~\cite{achouri2023a}. An example of such a structure is depicted in\figref{fig_fourier_planes}{(g)}. In this case, the expressions of the transmission and reflection coefficients are rather complicated and, for this reason, are shown in Eqs.~\textbf{(D24)-(D27)} of the SI. Note that if the susceptibilities $\chid{em}{zx}$ and $\chid{em}{xz}$ were chosen instead, then then transmission would be independent with respect to $k_y$ instead of $k_x$.

Referring to~Eqs.~\textbf{(D24)-(D27)}, it can be seen that the co- and cross-polarized reflection coefficients cancel out by setting the following conditions
\begin{align}
    \label{eq_xemyzxemze}
    \chid{em}{zy} = \chid{em}{yz},
\end{align}
which, as explained in previous sections, may be achieved by exploiting multiple Lorentzian resonances of the susceptibilities that overlap each other within a narrow frequency band. If the condition in~\eqqref{eq_xemyzxemze} is satisfied, then the TM co- and cross-polarized reflection coefficients in \textbf{(D24)-(D27)} are zero. The resulting scattering parameters reduce to
\begin{subequations}
\label{eq_chiemyzTR}
\begin{align}
T\@{co}(\theta, \phi) &= \frac{4 + k_y^2 \chi\@{em}^{{yz}^2}}{4 - k_y^2 \chi\@{em}^{{yz}^2}}, \quad R\@{co}(\theta, \phi)  = 0, \label{eq_chiemxyT} \\
T\@{cr}(\theta, \phi) &= \frac{4jk_y \chi\@{em}^{yz}}{4 - k_y^2 \chi\@{em}^{{yz}^2}}, \quad R\@{cr}(\theta, \phi)  = 0.
\end{align}
\end{subequations}
Performing the same analysis in the case of a TE polarized wave, leads to similar relations for the co- and cross-polarized transmission coefficients such that \mbox{$T\@{TE/TE}=T\@{TM/TM}$} and $T\@{TE/TM}=-T\@{TM/TE}$, respectively. Substituting these relations into the transmission matrix in the CP basis given in Eq.~\textbf{(D15)} leads to
\begin{equation}
\label{eq_Tcpchiral}
\te{T}\@{CP} =
T_{\text{co}}
\begin{bmatrix}
1 &0\\
0&1
\end{bmatrix} 
+ j T_{\text{cr}}
\begin{bmatrix}
1 &0\\
0&-1
\end{bmatrix}.
\end{equation}
This relation shows that such a metasurface exhibits circular birefringence/dichroism since the diagonal elements of $\te{T}\@{CP}$ are not equal when $T_{\text{cr}}\neq 0$. This is an interesting result since circular birefringence generally occurs in chiral media. However, our structure is \emph{not} chiral since it is reflection symmetric along both the $y$- and $z$-directions and one may, a priori, think that it should therefore not exhibit circular birefringence properties. Nevertheless, our structure is in fact pseudochiral due to its broken reflection symmetry along the $x$-direction~\cite{tretyakov2001,achouri2023a}. And, it turns out that a pseudochiral structure may still exhibit \emph{extrinsic} chiral responses if it is illuminated in the direction along which it is spatially symmetric~\cite{plum2009}. This extrinsic chiral effect is clearly apparent in~\eqqref{eq_chiemyzTR} and~\eqqref{eq_Tcpchiral} as it only exists if $k_y\neq 0$. In the case where $k_y=0$, then $T_\text{cr} = 0$ leading to identical transmission properties for LCP and RCP waves.

To illustrate the response of such a metasurface, we plot the transmission coefficients in the Fourier plane for a case where $\chid{em}{yz}=\chid{em}{zy}=8j/k$ in\figref{fig_fourier_planes}{(h,i)}. We clearly see that the transmission response of this metasurface is independent of the wavevector component $k_x$. As can be seen from\figref{fig_fourier_planes}{(h,i)}, there are two lines for which the amplitude of the cross-polarization transmission coefficient is maximized and equals unity. From~\eqqref{eq_chiemxyT}, we see that this occurs when $k_y = \pm 2/\text{Im}(\chid{em}{yz})$, which, in our case, corresponds to $k_y/k=\pm 1/4$. Interestingly, the phase of the cross-polarized transmission only takes binary values, which corresponds to $\pi$ when $k_y > 0$, and to 0 when $k_y < 0$. Importantly, this result demonstrates that it is theoretically possible to achieve unity amplitude extrinsic chiral effects with such a pseudochiral structure.

\section{\label{sec:conclusion} Conclusion}

We have derived and studied angular invariance conditions for both isolated particles and for metasurfaces. For isolated particles, achieving rotational invariance was associated with isotropic or bi-isotropic polarizabilities. It was also shown that such particles, while being rotationally invariant, are generally dependent on the direction of propagation of the illumination, which might a priori appear counter-intuitive.

For metasurfaces, we utilized the GSTCs framework to derive scattering coefficients, enabling us to determine conditions on the susceptibilities for achieving partial or complete angle-invariant responses. This framework was applied to various types of metasurfaces, including diagonal anisotropic, off-diagonal anisotropic, and bianisotropic configurations. In each case, we identified specific susceptibility conditions that provide angular independence for the reflection and transmission amplitude, phase, and the conversion of polarization. Several of these cases were also illustrated using full-wave simulated structures either in the microwave and optical regimes.

Interestingly, we see that a metasurface provides much more diversity in its ability to achieve different types of angle-invariant scattering responses compared to isolated particle even when considering the same modeling framework, i. e., dipolar responses modeled using bianisotropic effective material parameters.

Overall, this work demonstrates that: 1) nonlocality may not only be used to increase the angular dispersive response of a metasurface, as commonly believed, but also as a mean to completely remove all angular dependence; 2) spatial differentiation is possible in an angle-invariant fashion with unity amplitude; 3) fully-efficient polarization conversion based on an extrinsic chirality phenomenon is achievable in a partially angle-invariant fashion by exploiting pseudochiral responses.

This work paves the way for the implementation of angular filtering operations and other optical analog processing applications whose transfer function must remain independent of the direction of wave propagation or the physical orientation of the optical device itself.

\begin{acknowledgments}
We acknowledge funding from the Swiss National Science Foundation (project TMSGI2\_218392).
\end{acknowledgments}

\bibliography{apssamp}

\appendix
\setcounter{figure}{0}
\renewcommand{\thefigure}{S\arabic{figure}} 

\onecolumngrid 
\newpage
\section*{Supplementary Information}
\appendix

\renewcommand{\theequation}{A\arabic{equation}}
\renewcommand{\thefigure}{S\arabic{figure}}
\renewcommand{\thesubsection}{\thesection\arabic{subsection}}

\renewcommand{\theequation}{A\arabic{equation}}
\section{Angle Invariance and Wavevector-Dependence of an Isolated Particle}

The electric field scattered by a dipolar particle is given by
\begin{equation}
\label{eq_Esdip}
\ve{E}_\text{s} = \frac{G\@{0}}{\epsilon_0}\left[\ve{\hat{r}}\times(\ve{p}\times\ve{\hat{r}})+\frac{1}{c_0}\ve{m}\times\ve{\hat{r}}\right],
\end{equation}
where the dipole moments are
\begin{equation}
\begin{bmatrix}
    \ve{p}\\
    \ve{m}
\end{bmatrix}=
\begin{bmatrix}
    \epsilon_0\te{\alpha}\@{ee} & \frac{1}{c_0} \te{\alpha}\@{em}\\
    \frac{1}{\eta_0}\te{\alpha}\@{me} &  \te{\alpha}\@{mm}
\end{bmatrix}\cdot
\begin{bmatrix}
    \ve{E}\@{i}\\
    \ve{H}\@{i}
\end{bmatrix}.
\end{equation}
Assuming that the particle is bi-isotropic $(\te{\alpha}\rightarrow\alpha)$, we substitute the corresponding dipole moments into~\eqqref{eq_Esdip} to get
\begin{subequations}
\begin{align}
\ve{E}_\text{s}&= \frac{G\@{0}}{\epsilon_0}\Bigg[\ve{\hat{r}}\times\left(\epsilon_0\alpha\@{ee}\ve{E}\@{i} \times\ve{\hat{r}}+\frac{1}{c_0}\alpha\@{em}\ve{H}\@{i}\times\ve{\hat{r}}\right)+\frac{1}{c_0}\left(\alpha\@{mm}\ve{H}\@{i} +\frac{1}{\eta_0}\alpha\@{me}\ve{E}\@{i}\right)\times\ve{\hat{r}}\Bigg],\\
&= G\@{0}\left[\ve{\hat{r}}\times\left(\alpha\@{ee}\ve{E}\@{i} \times\ve{\hat{r}}+\eta_0\alpha\@{em}\ve{H}\@{i}\times\ve{\hat{r}}\right)+\left(\eta_0\alpha\@{mm}\ve{H}\@{i} +\alpha\@{me}\ve{E}\@{i}\right)\times\ve{\hat{r}}\right].
\end{align}
\end{subequations}
We now develop the cross-product to get only double cross-products using 
\begin{equation}
    \ve{E} = \eta_0 \ve{H}\times\ve{\hat{k}} \quad\text{and}\quad     \ve{H} =  \ve{\hat{k}}\times\ve{E}/\eta_0,
\end{equation}
we have
\begin{equation}
\begin{split}
\ve{E}_\text{s}&=G\@{0}\Bigg[\alpha\@{ee}\ve{\hat{r}}\times(\ve{E}\@{i} \times\ve{\hat{r}})+\eta_0\alpha\@{em}\ve{\hat{r}}\times(\ve{H}\@{i}\times\ve{\hat{r}})\\
&\qquad\qquad\qquad+\alpha\@{mm}(\ve{\hat{k}}\times\ve{E}\@{i})\times\ve{\hat{r}} +\eta_0\alpha\@{me}( \ve{H}\@{i}\times\ve{\hat{k}})\times\ve{\hat{r}}\Bigg].
\end{split}
\end{equation}
Flipping the last two double cross-products, we have
\begin{equation}
\label{eq_Es1}
\begin{split}
\ve{E}_\text{s}&=G\@{0}\Big[\alpha\@{ee}\ve{\hat{r}}\times(\ve{E}\@{i} \times\ve{\hat{r}})+\eta_0\alpha\@{em}\ve{\hat{r}}\times(\ve{H}\@{i}\times\ve{\hat{r}})\\
&\qquad\qquad\qquad-\alpha\@{mm}\ve{\hat{r}}\times(\ve{\hat{k}}\times\ve{E}\@{i}) -\eta_0\alpha\@{me}\ve{\hat{r}}\times(\ve{H}\@{i}\times\ve{\hat{k}})\Big].
\end{split}
\end{equation}
Considering the relation
\begin{equation}
    \ve{A}\times(\ve{B}\times\ve{C}) = (\ve{A}\cdot\ve{C})\ve{B}-\ve{C}(\ve{A}\cdot\ve{B}),
\end{equation}
we transform~\eqqref{eq_Es1} into
\begin{equation}
\label{eq_es2}
\begin{split}
\ve{E}_\text{s}&=G\@{0}\Big[\alpha\@{ee}[\ve{E}\@{i}-\ve{\hat{r}}(\ve{\hat{r}}\cdot\ve{E}\@{i})]-\alpha\@{mm}[(\ve{\hat{r}}\cdot\ve{E}\@{i})\ve{\hat{k}}-(\ve{\hat{r}}\cdot\ve{\hat{k}})\ve{E}\@{i}]\\
&\qquad\qquad\qquad +\eta_0\alpha\@{em}[\ve{H}\@{i}-\ve{\hat{r}}(\ve{\hat{r}}\cdot\ve{H}\@{i})]-\eta_0\alpha\@{me}[(\ve{\hat{r}}\cdot\ve{\hat{k}})\ve{H}\@{i}-(\ve{\hat{r}}\cdot\ve{H}\@{i})\ve{\hat{k}}]\Big].
\end{split}
\end{equation}
We now factorize $\ve{E}\@{i}$ and $\ve{H}\@{i}$, which yields
\begin{equation}
\begin{split}   
\ve{E}_\text{s}&=G\@{0}\Big\{\left[\alpha\@{ee}\left(\te{I}-\ve{\hat{r}}\ve{\hat{r}}\right)-\alpha\@{mm}\left[\ve{\hat{k}}\ve{\hat{r}}-\left(\ve{\hat{r}}\cdot\ve{\hat{k}}\right)\te{I}\right]\right]\cdot\ve{E}\@{i} \\
&\qquad\qquad\qquad\qquad+\eta_0\left[\alpha\@{em}\left(\te{I}-\ve{\hat{r}}\ve{\hat{r}}\right)+\alpha\@{me}\left[\ve{\hat{k}}\ve{\hat{r}}-\left(\ve{\hat{r}}\cdot\ve{\hat{k}}\right)\te{I}\right]\right]\cdot\ve{H}\@{i}\Big\}.
\end{split}
\end{equation}
Expressing $\ve{H}\@{i}$ in terms of $\ve{E}\@{i}$ gives
\begin{equation}
\begin{split}   
\ve{E}_\text{s}&=G\@{0}\Big\{\left[\alpha\@{ee}\left(\te{I}-\ve{\hat{r}}\ve{\hat{r}}\right)-\alpha\@{mm}\left[\ve{\hat{k}}\ve{\hat{r}}-\left(\ve{\hat{r}}\cdot\ve{\hat{k}}\right)\te{I}\right]\right]\cdot\ve{E}\@{i} \\
&\qquad\qquad\qquad\qquad+\left[\alpha\@{em}\left(\te{I}-\ve{\hat{r}}\ve{\hat{r}}\right)+\alpha\@{me}\left[\ve{\hat{k}}\ve{\hat{r}}-\left(\ve{\hat{r}}\cdot\ve{\hat{k}}\right)\te{I}\right]\right]\cdot\ve{\hat{k}}\times\ve{E}\@{i}\Big\}.
\end{split}
\end{equation}
Factorizing again $\ve{E}\@{i}$ leads to
\begin{equation}
\begin{split}   
\ve{E}_\text{s}&=G\@{0}\Big\{\alpha\@{ee}\left(\te{I}-\ve{\hat{r}}\ve{\hat{r}}\right)-\alpha\@{mm}\left[\ve{\hat{k}}\ve{\hat{r}}-\left(\ve{\hat{r}}\cdot\ve{\hat{k}}\right)\te{I}\right]\\
&\qquad\qquad\qquad\qquad+\left[\alpha\@{em}\left(\te{I}-\ve{\hat{r}}\ve{\hat{r}}\right)+\alpha\@{me}\left[\ve{\hat{k}}\ve{\hat{r}}-\left(\ve{\hat{r}}\cdot\ve{\hat{k}}\right)\te{I}\right]\right]\cdot\te{k}\Big\}\cdot\ve{E}\@{i},
\end{split}
\end{equation}
where $\te{k} = \ve{\hat{k}}\times\te{I}$. We may thus express the scattered electric field as
\begin{equation} 
\ve{E}_\text{s}= G\@{0}\te{G}(\ve{k})\cdot\ve{E}\@{i},
\end{equation}
where $\te{G}(\ve{k})$ is a dyadic Green function defined by
\begin{equation}
    \te{G}(\ve{k}) = \te{G}_1(\ve{k}) + \te{G}_2(\ve{k}),
\end{equation}
with
\begin{align}
\te{G}_1(\ve{k})&=\alpha\@{ee}\left(\te{I}-\ve{\hat{r}}\ve{\hat{r}}\right)-\alpha\@{mm}\left[\ve{\hat{k}}\ve{\hat{r}}-\left(\ve{\hat{r}}\cdot\ve{\hat{k}}\right)\te{I}\right],\\
\te{G}_2(\ve{k}) &=\left[\alpha\@{em}\left(\te{I}-\ve{\hat{r}}\ve{\hat{r}}\right)+\alpha\@{me}\left[\ve{\hat{k}}\ve{\hat{r}}-\left(\ve{\hat{r}}\cdot\ve{\hat{k}}\right)\te{I}\right]\right]\cdot\te{k}.
\end{align}
Under a spatial transformation $\te{\Lambda}$, the dyadic $\te{G}(\ve{k})$ transforms in the new basis into $\te{G}'(\ve{k})$, which is given by
\begin{equation}
    \te{G}'(\ve{k}) = \te{\Lambda}\cdot\te{G}(\ve{k})\cdot\te{\Lambda}^{-1}.
\end{equation}
Since $\te{\Lambda}^{-1} = \te{\Lambda}^{T}$ for rotation and reflection operations (defined by orthogonal matrices), we have that
\begin{equation}
    \te{G}'(\ve{k}) =  \te{G}_1(\ve{k}) + \text{det}\left(\te{\Lambda}\right)\te{G}_2(\ve{k}),
\end{equation}
where $\text{det}\left(\te{\Lambda}\right) = 1$ for rotations and $\text{det}\left(\te{\Lambda}\right)=-1$ for reflections. It follows that $\te{G}_1(\ve{k})$ is invariant under both rotations and reflections, whereas $\te{G}_2(\ve{k})$ is invariant only under rotations.

\newpage
\renewcommand{\theequation}{B\arabic{equation}}
\section{\label{sec_theo_background}Theoretical Background}
This section of the supplementary material would give some more detail about the convention used to derive the transmission equation of the main text. The boundary condition can be expressed as 
\begin{align}
\hat{z} \times \Delta \ve{H} &= j \omega \ve{P}_\parallel - \hat{z} \times \nabla_\parallel M_z \label{eq:A1} \\
\hat{z} \times \Delta \ve{E} &= -j \omega \mu_0 \ve{M}_\parallel - \hat{z} \times \nabla_\parallel \left(\frac{P_z}{\epsilon_0}\right) \label{eq:A2},
\end{align}
Where we can define the difference such as 
\begin{align}
     \Delta \ve{H} &= \ve{H}^{z=0^+}_\parallel-\ve{H}^{z=0^-}_\parallel \label{eq:A3} \\
     \Delta \ve{E} &= \ve{E}^{z=0^+}_\parallel-\ve{E}^{z=0^-}_\parallel \label{eq:A4},
\end{align}
and the constitutive equations
\begin{align}
\ve{P} &= \epsilon_0 \te{\chi}_\text{ee} \cdot \ve{E}_\text{av} + \frac{1}{c_0} \te{\chi}_\text{em} \cdot \ve{H}_\text{av} \label{eq:A5} \\
\ve{M} &= \te{\chi}_\text{mm} \cdot \ve{H}_\text{av} + \frac{1}{\eta_0} \te{\chi}_{me} \cdot \ve{E}_\text{av} \label{eq:A6}.
\end{align}
The parallel component of the polarization is denoted as $\ve{P}_\parallel$ and it defines the component of the surface (perpendicular to the normal of the surface). So the parallel gradit is equal to $\nabla_\parallel = \left( \frac{\partial}{\partial x}, \frac{\partial}{\partial y} \right)
$. The average fields can be defined as
\begin{align}
    \ve{E}_\text{av} &= \frac{1}{2}(\ve{E}_\text{1} + \ve{E}_\text{2}) \\
    \ve{H}_\text{av} &= \frac{1}{2}(\ve{H}_\text{1} + \ve{H}_\text{2}).
\end{align}

Where $\ve{E}_\text{1}$, $\ve{H}_\text{1}$ and $\ve{E}_\text{2}$, $\ve{H}_\text{2}$ are the fields calculated at the interface for $z = 0^+$ and $z = 0^-$ respectively. Now let us define the incident, reflected and transmitted magnetic fields for a TM polarization
\begin{align}
    \ve{H}_\text{i} &= -H_0e^{-j(k_{x} x-k_{z} z)}\hat{y} \\
    \ve{H}_\text{r,co} &= R_\text{co}H_0e^{-j(k_{x} x+k_{z} z)}\hat{y} \\
    \ve{H}_\text{t,co} &= -T_\text{co}H_0e^{-j(k_{x} x-k_{z} z)}\hat{y},
\end{align}
with $H_0$ as the amplitude of the magnetic field for the incident wave. The relation between the magnetic and electric field is defined by
\begin{align}
    \ve{E} &= \frac{\eta_\text{0}}{k_\text{0}} \ve{H} \times \ve{k}.
\end{align}
Thus, we could obtain the electric field as 
\begin{align}
    \ve{E}_\text{i} &= \frac{\eta_\text{0}}{k_\text{0}}H_0e^{-j(k_{x} x - k_{z} z)} \left[ k_{x}\hat{z} + k_{z}\hat{x} \right] \\
    \ve{E}_\text{r,co} &= -R_\text{co}H_0\frac{\eta_\text{0}}{k_\text{0}}e^{-j(k_{x} x + k_{z} z)}\left[ k_{x}\hat{z} - k_{z}\hat{x} \right] \\
    \ve{E}_\text{t,co} &= T_\text{co}H_0\frac{\eta_\text{0}}{k_\text{0}}e^{-j(k_{x} x - k_{z} z)}\left[ k_{x}\hat{z} + k_{z}\hat{x} \right] 
\end{align}
We should also consider the cross polarization fields that are defined as
\begin{align}
    \ve{E}_\text{r,cr} &= R_\text{cr}\eta_\text{0}H_0e^{-j(k_{x} x + k_{z} z)}\hat{y} \\
    \ve{E}_\text{t,cr} &= T_\text{cr}\eta_\text{0}H_0e^{-j(k_{x} x - k_{z} z)}\hat{y} \\
    \ve{H}_\text{r,cr} &= \frac{R_\text{cr}}{k_\text{0}}H_0e^{-j(k_{x} x + k_{z} z)}\left[ k_{x}\hat{z} - k_{z}\hat{x} \right] \\
    \ve{H}_\text{t,cr} &= \frac{T_\text{cr}}{k_\text{0}}H_0e^{-j(k_{x} x - k_{z} z)}\left[ k_{x}\hat{z} + k_{z}\hat{x} \right]
\end{align}
Thus, by gathering all the equations, we can derive the transmission and reflection coefficient for both cross- and co-polarization.

\newpage
\renewcommand{\theequation}{C\arabic{equation}}
\section{\label{sec_anis_met}Anisotropic metasurface}
Anisotropic metasurfaces are the ones that does not present any coupling between the magnetic and electric fields.
\begin{align}
    \te{\chi}_\text{ee} = 
    \begin{pmatrix}
        \chid{ee}{xx}&  
        \chid{ee}{xy}&
        \chid{ee}{xz}\\
        \chid{ee}{yx}&
        \chid{ee}{yy}&
        \chid{ee}{yz}\\
        \chid{ee}{zx}&
        \chid{ee}{zy}&
        \chid{ee}{zz}
    \end{pmatrix}, \quad
    \te{\chi}_\text{mm} = 
    \begin{pmatrix}
        \chid{mm}{xx}&  
        \chid{mm}{xy}&
        \chid{mm}{xz}\\
        \chid{mm}{yx}&
        \chid{mm}{yy}&
        \chid{mm}{yz}\\
        \chid{mm}{zx}&
        \chid{mm}{zy}&
        \chid{mm}{zz}
    \end{pmatrix}, \quad
    \te{\chi}_\text{em} = 0, \quad
    \te{\chi}_\text{me} = 0.
\end{align}
In the following sections and also in the main paper, not all the susceptibilities are considered due to the symmetry of the structure but also due to the polarization of the incoming wave.
\subsection{Diagonal Anisotropic Metasurface}
Here, we start to consider a very simple case, the diagonal anisotropic metasurface that only has the diagonal elements that are non-zero.
\begin{align}
    \te{\chi}_\text{ee} = 
    \begin{pmatrix}
        \chid{ee}{xx}&  
        0&
        0\\
        0&
        \chid{ee}{yy}&
        0\\
        0&
        0&
        \chid{ee}{zz}
    \end{pmatrix}, \quad
    \te{\chi}_\text{mm} = 
    \begin{pmatrix}
        \chid{mm}{xx}&  
        0&
        0\\
        0&
        \chid{mm}{yy}&
        0\\
        0&
        0&
        \chid{mm}{zz}
    \end{pmatrix}, \quad
    \te{\chi}_\text{em} = 0, \quad
    \te{\chi}_\text{me} = 0.
\end{align}
Depending on the polarization of the incoming wave, some element will not be excited. For example for TM polarization, the resulting tensors become,
\begin{align}
    \te{\chi}_\text{ee} = 
    \begin{pmatrix}
        \chid{ee}{xx}&  
        0&
        0\\
        0&
        0&
        0\\
        0&
        0&
        \chid{ee}{zz}
    \end{pmatrix}, \quad
    \te{\chi}_\text{mm} = 
    \begin{pmatrix}
        0&  
        0&
        0\\
        0&
        \chid{mm}{yy}&
        0\\
        0&
        0&
        0
    \end{pmatrix}, \quad
    \te{\chi}_\text{em} = 0, \quad
    \te{\chi}_\text{me} = 0.
\end{align}
And for TE polarized light,
\begin{align}
    \te{\chi}_\text{ee} = 
    \begin{pmatrix}
        0&  
        0&
        0\\
        0&
        \chid{ee}{yy}&
        0\\
        0&
        0&
        0
    \end{pmatrix}, \quad
    \te{\chi}_\text{mm} = 
    \begin{pmatrix}
        \chid{mm}{xx}&  
        0&
        0\\
        0&
        0&
        0\\
        0&
        0&
        \chid{mm}{zz}
    \end{pmatrix}, \quad
    \te{\chi}_\text{em} = 0, \quad
    \te{\chi}_\text{me} = 0.
\end{align}
For the case of a TM polarized case, we only consider the susceptibilities $\chi_\text{ee}^{xx}$, $\chi_\text{ee}^{zz}$ and $\chi_\text{mm}^{yy}$. Using the equation built previously, we can obtain the transmission and reflection with the same procedure, in the case the waves propagate in the $xz$ plane ($k_y=0$):
\begin{align}
R_{\text{co}}(\theta) &= \frac{2 \left(k_{x}^2 \chi_{\text{ee}}^{{zz}} + k^2 \chi_{\text{mm}}^{{yy}} - k_{z}^2 \chi_{\text{ee}}^{{xx}}\right)}{\left(k_{z} \chi_{\text{ee}}^{{xx}} - 2j\right) \left[2 k_{z} + j \left(k_{x}^2 \chi_{\text{ee}}^{{zz}} + k^2 \chi_{\text{mm}}^{{yy}}\right)\right]} \\
T_{\text{co}}(\theta) &= -\frac{j k_{z} \left(4 + k_{x}^2 \chi_{\text{ee}}^{{xx}} \chi_{\text{ee}}^{{zz}} + k^2 \chi_{\text{ee}}^{{xx}} \chi_{\text{mm}}^{{yy}}\right)}{\left(k_{z} \chi_{\text{ee}}^{{xx}} - 2j\right) \left[2 k_{z} + j \left(k_{x}^2 \chi_{\text{ee}}^{{zz}} + k^2 \chi_{\text{mm}}^{{yy}}\right)\right]},
\label{eq_1}
\end{align}
with $k_x=k \sin\theta$ and $k_z=k \cos\theta$. From these equations, the conditions for having an angular invariant are visible. By selecting properly the conditions explained in the main text we can achieve different angular invariance.

Let's take the equation above for which the co-reflection and co-transmission can be separated into the real and imaginary part. 

\begin{align}
R_{\text{co}}(\theta) &= R_{\text{co},\mathbb{R}}(\theta) + j R_{\text{co},\mathbb{I}}(\theta) \\
T_{\text{co}}(\theta) &= T_{\text{co},\mathbb{R}}(\theta) + j T_{\text{co},\mathbb{I}}(\theta)
\end{align}

Using Mathematica, we can find condition that cancel the real part or imaginary part of the transmission and reflection. The solutions are double for each case.

\begin{align}
\Re(R_\text{co}(\theta)) = 0  \Longleftrightarrow  
\begin{cases}
\chi_{\text{mm,1}}^{{yy}}(\theta) = -\frac{1}{k^2} \left(k_z^2 \chi_{\text{ee}}^{{xx}} + k_x^2 \chi_{\text{ee}}^{{zz}}\right)\\
\chi_{\text{mm,2}}^{{yy}}(\theta) = \frac{1}{k^2} \left(k_z^2 \chi_{\text{ee}}^{{xx}} - k_x^2 \chi_{\text{ee}}^{{zz}}\right)
\end{cases}    
\end{align}
\begin{align}
\Im(R_\text{co}(\theta)) = 0  \Longleftrightarrow  
\begin{cases}
\chi_{\text{mm,1}}^{{yy}}(\theta) = \frac{1}{\chi_{\text{ee}}^{{xx}}k^2} \left(4 - k_x^2 \chi_{\text{ee}}^{{xx}}\chi_{\text{ee}}^{{zz}}\right) \\
\chi_{\text{mm,2}}^{{yy}}(\theta) = \frac{1}{k^2} \left(k_z^2 \chi_{\text{ee}}^{{xx}} - k_x^2 \chi_{\text{ee}}^{{zz}}\right)
\end{cases}    
\end{align}
\begin{align}
\Re(T_\text{co}(\theta)) = 0  \Longleftrightarrow  
\begin{cases}
\chi_{\text{mm,1}}^{{yy}}(\theta) = \frac{1}{\chi_{\text{ee}}^{{xx}}k^2} \left(4 - k_x^2 \chi_{\text{ee}}^{{xx}}\chi_{\text{ee}}^{{zz}}\right) \\
\chi_{\text{mm,2}}^{{yy}}(\theta) = -\frac{1}{\chi_{\text{ee}}^{{xx}}k^2} \left(4 + k_x^2 \chi_{\text{ee}}^{{xx}}\chi_{\text{ee}}^{{zz}}\right)
\end{cases}    
\end{align}
\begin{align}
\Im(T_\text{co}(\theta))= 0  \Longleftrightarrow  
\begin{cases}
\chi_{\text{mm,1}}^{{yy}}(\theta) = -\frac{1}{\chi_{\text{ee}}^{{xx}}k^2} \left(4 + k_x^2 \chi_{\text{ee}}^{{xx}}\chi_{\text{ee}}^{{zz}}\right) \\
\chi_{\text{mm,2}}^{{yy}}(\theta) = -\frac{1}{k^2} \left(k_z^2 \chi_{\text{ee}}^{{xx}} + k_x^2 \chi_{\text{ee}}^{{zz}}\right)
\end{cases}    
\end{align}

We observe that, for certain solutions, both the real and imaginary parts of the transmission or reflection simultaneously become zero, resulting in complete transmission or reflection. It is also possible to select a specific solution to independently nullify either the real or imaginary component of either reflection or transmission. This so-called Kerker condition dictates that the imaginary part of the reflection and the real part of the transmission must vanish. Consequently, this implies that the reflection is purely real, and the transmission is purely imaginary, leading to a constant phase shift.

We can clearly notice that for $\chi_{\text{ee}}^{zz} = 0$ and $\chi_{\text{mm}}^{{yy}} = \chi_{\text{ee}}^{{zz}}$, then we can find out the Kerker condition mentionned in the main text. The specific conditions of the susceptibilities could also be determined in order to remove the angular dependencies of the solutions.

\subsection{Off-Diagonal Anisotropic Metasurfaces}
Off-diagonal anisotropic metasurfaces have the diagonal elements of the susceptibilities matrix null, as it follows,
\begin{align}
    \te{\chi}_\text{ee} = 
    \begin{pmatrix}
        0&  
        \chid{ee}{xy}&
        \chid{ee}{xz}\\
        \chid{ee}{yx}&
        0&
        \chid{ee}{yz}\\
        \chid{ee}{zx}&
        \chid{ee}{zy}&
        0
    \end{pmatrix}, \quad
    \te{\chi}_\text{mm} = 
    \begin{pmatrix}
        0&  
        \chid{mm}{xy}&
        \chid{mm}{xz}\\
        \chid{mm}{yx}&
        0&
        \chid{mm}{yz}\\
        \chid{mm}{zx}&
        \chid{mm}{zy}&
        0
    \end{pmatrix}, \quad
    \te{\chi}_\text{em} = 0, \quad
    \te{\chi}_\text{me} = 0.
\end{align}
Again these tensors are different when illuminating with a TM or TE polarized beam.
\subsubsection{Purely Off-Diagonal Case with Normal Polarization}
If now we consider that both susceptibilities $\chid{ee}{yz}$ and $\chid{mm}{yz}$ exist, thus the tensors become,
\begin{align}
    \te{\chi}_\text{ee} = 
    \begin{pmatrix}
        0&  
        0&
        0\\
        0&
        0&
        \chid{ee}{yz}\\
        0&
        \chid{ee}{zy}&
        0
    \end{pmatrix}, \quad
    \te{\chi}_\text{mm} = 
    \begin{pmatrix}
        0&  
        0&
        0\\
        0&
        0&
        \chid{mm}{yz}\\
        0&
        \chid{mm}{zy}&
        0
    \end{pmatrix}, \quad
    \te{\chi}_\text{em} = 0, \quad
    \te{\chi}_\text{me} = 0.
\end{align}
The coefficients for TE illumination become complicated expressions but their proportionalities are as follow,

\begin{align}
R\@{co}(\theta, \phi) &= -4 k^2 k_x^2 (\chid{ee}{yz} - \chid{mm}{yz})^2 /D(\theta, \phi), \\ 
T\@{co}(\theta, \phi) &= - k_z^2 \left( k_y \chid{mm}{yz}-2j\right)^2 \left(4 + k_y^2 \chid{ee}{yz^2}\right)/D(\theta, \phi), \\ 
R\@{cr}(\theta, \phi) &= -2j k k_x k_z \left( k_y \chid{ee}{yz}+2j\right) (\chid{ee}{yz} - \chid{mm}{yz}) \left(k_y \chid{mm}{yz} -2j\right)/D(\theta, \phi), \\ 
T\@{cr}(\theta, \phi) &= -2j k k_x k_z \left( k_y \chid{ee}{yz}-2j\right) (\chid{ee}{yz} - \chid{mm}{yz}) \left(k_y \chid{mm}{yz}-2j\right)/D(\theta, \phi),
\end{align}
where $k_t^2=k_x^2+k_y^2$ and $D(\theta, \phi)$ is the denominator and is expressed such as,
\begin{equation}
D(\theta, \phi) =k_z^2 \left(16 + k_y^4 \chid{ee}{yz^2} \chid{mm}{yz^2} \right) + 4 k_t^2 \left(k^2 - k_y^2 \right) \left( \chid{ee}{yz^2} + \chid{mm}{yz^2}\right) -8 k^2 k_x^2 \chid{ee}{yz} \chid{mm}{yz}.
\end{equation}
It can be clearly seen that we can cancel out the co- and cross- polarized reflection but also the cross polarized transmission by setting
\begin{align}
    \chid{ee}{yz} = \chid{mm}{yz}.
\end{align}
This condition reduces the coefficients to,
\begin{align}
R\@{co}(\theta, \phi) &= 0, \\
T\@{co}(\theta, \phi) &= \frac{2j - k_y \chid{mm}{yz}}{2j + k_y \chid{mm}{yz}}, \\
R\@{cr}(\theta, \phi) &= 0, \\
T\@{cr}(\theta, \phi) &= 0.
\end{align}
These equations show clearly that the amplitude in transmission is always 1, but the phase varies with $k_y$.
We can also consider the case with $\chid{mm}{yz} = 0$, thus the only remaining susceptibility is $\chid{ee}{yz}$. The susceptibility tensors become,
\begin{align}
    \te{\chi}_\text{ee} = 
    \begin{pmatrix}
        0&  
        0&
        0\\
        0&
        0&
        \chid{ee}{yz}\\
        0&
        \chid{ee}{zy}&
        0
    \end{pmatrix}, \quad
    \te{\chi}_\text{mm} = 0, \quad
    \te{\chi}_\text{em} = 0, \quad
    \te{\chi}_\text{me} = 0.
\end{align}
The coefficients are given by,
\begin{align}
R\@{co}(\theta, \phi) &= -\frac{k^2 k_x^2 \chid{ee}{yz^2}}{
k^2 k_x^2 \chid{ee}{yz^2} + k_z^2 \left(4 + k_y^2 \chid{ee}{yz^2}\right)} \\
   T_{\text{co}}(\theta, \phi) &= \frac{k_z^2(4+k_y^2\chid{ee}{{yz}^2})}{4k_z^2+(k^2k_x^2+k_y^2k_z^2)\chid{ee}{{yz}^2}}. \\
R\@{cr}(\theta, \phi) &= -\frac{k k_x k_z \chid{ee}{yz} \left(k_y \chid{ee}{yz}-2j\right)}{
k^2 k_x^2 \chid{ee}{yz^2} + k_z^2 \left(4 + k_y^2 \chid{ee}{yz^2}\right)} \\
T\@{cr}(\theta, \phi) &=- \frac{k k_x k_z \chid{ee}{yz} \left(k_y \chid{ee}{yz}-2j\right)}{
k^2 k_x^2 \chid{ee}{yz^2} + k_z^2 \left(4 + k_y^2 \chid{ee}{yz^2}\right)}
\end{align}
It can be clearly seen that for $k_x = 0$, the transmission coefficient reduces to $|T\@{co}(\theta, \phi)| = 1$, as explained in the main text.
\subsubsection{Partial Gyrotropic Angular Invariance}
\begin{align}
    \te{\chi}_\text{ee} = 
    \begin{pmatrix}
        0&  
        \chid{ee}{xy}&
        0\\
        \chid{ee}{yx}&
        0&
        0\\
        0&
        0&
        0
    \end{pmatrix}, \quad
    \te{\chi}_\text{mm} = 
    \begin{pmatrix}
        0&  
        \chid{mm}{xy}&
        0\\
        \chid{mm}{yx}&
        0&
        0\\
        0&
        0&
        0
    \end{pmatrix}, \quad
    \te{\chi}_\text{em} = 0, \quad
    \te{\chi}_\text{me} = 0.
\end{align}
The coefficients become very complex for illumination coming from an arbitrary direction, however the scattering coefficients in the $xz$ ($k_y=0$) and the $yz$ ($k_x=0$) propagation planes could be written as it follows,
\begin{align}
R\@{co} &= \frac{4 k^2 \left(-\chid{ee}{xy^2} + \chid{mm}{xy^2}\right)}{\left(4 + k^2 \chid{ee}{xy^2}\right)\left(4 + k^2 \chid{mm}{xy^2}\right)}, \\ 
T\@{co} &= \frac{(4 + k^2\chid{ee}{xy}\chid{mm}{xy})(4 - k^2\chid{ee}{xy}\chid{mm}{xy})}{\left(4 + k^2 \chid{ee}{xy^2}\right)\left(4 + k^2 \chid{mm}{xy^2}\right)}, \\ 
R\@{cr} &= \mp\frac{2j k \left(\chid{ee}{xy} + \chid{mm}{xy}\right) \left(4 + k^2 \chid{ee}{xy} \chid{mm}{xy}\right)}{\left(4 + k^2 \chid{ee}{xy^2}\right)\left(4 + k^2 \chid{mm}{xy^2}\right)}, \\ 
T\@{cr} &= \pm\frac{2j k \left(\chid{ee}{xy} - \chid{mm}{xy}\right) \left(-4 + k^2 \chid{ee}{xy} \chid{mm}{xy}\right)}{\left(4 + k^2 \chid{ee}{xy^2}\right)\left(4 + k^2 \chid{mm}{xy^2}\right)}.
\end{align}
where the sign for the cross-polarized coefficients depends on the propagation plane, using the top sign for the case with $k_y=0$ and the bottom one for $k_x=0$. It can be observed that for,
\begin{align}
    \chid{ee}{xy} = \chid{mm}{xy}
\end{align}
we can obtain,
\begin{align}
R\@{co} &= 0, \\ 
T\@{co} &= \frac{4 - k^2 \chid{ee}{xy^2}}{4 + k^2 \chid{ee}{xy^2}}, \\ 
R\@{cr} &= \mp \frac{4j k \chid{ee}{xy}}{4 + k^2 \chid{ee}{xy^2}}, \\ 
T\@{cr} &= 0.
\end{align}
Another interesting case is if the following condition is satisfied,
\begin{align}
    \chid{ee}{xy} = -\chid{mm}{xy},
\end{align}
then,
\begin{align}
R\@{co} &= 0, \\ 
T\@{co} &= \frac{4 - k^2 \chid{ee}{xy^2}}{4 + k^2 \chid{ee}{xy^2}}, \\ 
R\@{cr} &= 0, \\ 
T\@{cr} &= \mp \frac{4j k \chid{ee}{xy}}{4 + k^2 \chid{ee}{xy^2}}.
\end{align}
For the particular case where $\chid{ee}{xy} = -\chid{mm}{xy} = \frac{2}{k}$,
we can obtain,
\begin{align}
R\@{co} &= 0, \\ 
T\@{co} &= 0 \\ 
R\@{cr} &= 0, \\ 
T\@{cr} &= \mp j.
\end{align}
 In the case $\chid{ee}{xy} = -\chid{mm}{xy} = -\frac{2}{k}$ is chosen instead, $T\@{cr}$ will have an additional $\pi$-shift in the form $T\@{cr} = \pm j$, where the top sign corresponds to propagation in the $xz$ plane, while the bottom describes propagation in the $yz$ plane instead. This means that for this special case we can obtain an angular invariance in phase of $\mp \pi / 2 $.

\newpage
\renewcommand{\theequation}{D\arabic{equation}}
\section{\label{sec_bianis_met}Bianisotropic metasurface}
Bianisotropic metasurface have all the possible tensors, such as
\begin{align}
    \te{\chi}_\text{ee} &= 
    \begin{pmatrix}
        \chid{ee}{xx}&  
        \chid{ee}{xy}&
        \chid{ee}{xz}\\
        \chid{ee}{yx}&
        \chid{ee}{yy}&
        \chid{ee}{yz}\\
        \chid{ee}{zx}&
        \chid{ee}{zy}&
        \chid{ee}{zz}
    \end{pmatrix}, \quad
    \te{\chi}_\text{mm} = 
    \begin{pmatrix}
        \chid{mm}{xx}&  
        \chid{mm}{xy}&
        \chid{mm}{xz}\\
        \chid{mm}{yx}&
        \chid{mm}{yy}&
        \chid{mm}{yz}\\
        \chid{mm}{zx}&
        \chid{mm}{zy}&
        \chid{mm}{zz}
    \end{pmatrix}, \\
    \te{\chi}_\text{em} &=
    \begin{pmatrix}
        \chid{em}{xx}&  
        \chid{em}{xy}&
        \chid{em}{xz}\\
        \chid{em}{yx}&
        \chid{em}{yy}&
        \chid{em}{yz}\\
        \chid{em}{zx}&
        \chid{em}{zy}&
        \chid{em}{zz}
    \end{pmatrix}, \quad
    \te{\chi}_\text{me} = 
    \begin{pmatrix}
        \chid{me}{xx}&  
        \chid{me}{xy}&
        \chid{me}{xz}\\
        \chid{me}{yx}&
        \chid{me}{yy}&
        \chid{me}{yz}\\
        \chid{me}{zx}&
        \chid{me}{zy}&
        \chid{me}{zz}
    \end{pmatrix}.
\end{align}
In this case, the electric-to-magnetic and magnetic-to-electric coupling is possible.
\subsection{Complete Nongyrotropic Angular Invariance}
In this section we only consider the susceptibilities $\chid{ee}{xx}$, $\chid{mm}{yy}$ and $\chid{em}{xy}$.
\begin{align}
    \te{\chi}_\text{ee} &= 
    \begin{pmatrix}
        \chid{ee}{xx}&  
        0&
        0\\
        0&
        0&
        0\\
        0&
        0&
        0
    \end{pmatrix}, \quad
    \te{\chi}_\text{mm} = 
    \begin{pmatrix}
        0&  
        0&
        0\\
        0&
        \chid{mm}{yy}&
        0\\
        0&
        0&
        0
    \end{pmatrix}, \\
    \te{\chi}_\text{em} &=
    \begin{pmatrix}
        0&  
        \chid{em}{xy}&
        0\\
        0&
        0&
        0\\
        0&
        0&
        0
    \end{pmatrix}, \quad
    \te{\chi}_\text{me} = 
    \begin{pmatrix}
        0&  
        0&
        0\\
        \chid{me}{yx}&
        0&
        0\\
        0&
        0&
        0
    \end{pmatrix}.
\end{align}
Thus, the coefficients obtained with these susceptibilities are as it follows,
\begin{subequations}
\begin{align}
R_{\text{co}}(\theta) &= \frac{- k_{z}^2 \chi_{\text{ee}}^{{xx}} \mp 2 k k_{z} \chi_{\text{em}}^{xy} + k^2 \chi_{\text{mm}}^{{yy}}}{k_{z}^2 \chi_{\text{ee}}^{{xx}} + k^2 \chi_{\text{mm}}^{{yy}} + j k_{z} \left[ \frac{k^2}{2} (\chi_{\text{em}}^{xy^2} + \chi_{\text{ee}}^{{xx}} \chi_{\text{mm}}^{{yy}})-2\right]} \label{eq:B1}, \\
T_{\text{co}}(\theta) &= \frac{-j k_{z} \left[\frac{k^2}{2} (\chi_{\text{em}}^{xy^2} + \chi_{\text{ee}}^{{xx}} \chi_{\text{mm}}^{{yy}})+2\right]}{k_{z}^2 \chi_{\text{ee}}^{{xx}} + k^2 \chi_{\text{mm}}^{{yy}} + j k_{z} \left[ \frac{k^2}{2} (\chi_{\text{em}}^{xy^2} + \chi_{\text{ee}}^{{xx}} \chi_{\text{mm}}^{{yy}})-2\right]}, \\
R\@{cr}(\theta) &= 0, \\
T\@{cr}(\theta) &= 0.
\label{eq:B2}
\end{align}
\end{subequations}
The minus and plus signs are for the forward and backward incidence field. Going further, we can assume that 
\begin{subequations}
\begin{align}
    \chi\@{ee}^{xx} = \chi\@{mm}^{yy} = 0, \\
    \chi\@{em}^{xy} \neq 0,
\end{align}
\end{subequations}
leading to,
\begin{subequations}
\label{eq:chiemxyTR}
\begin{align}
    T\@{co}(\theta) &= \frac{4 + k^2 \chi\@{em}^{{xy}^2}}{4 - k^2 \chi\@{em}^{{xy}^2}}, \label{eq:chiemxyTsi} \\
    R\@{co}(\theta) &= \pm\frac{4jk \chi\@{em}^{xy}}{4 - k^2 \chi\@{em}^{{xy}^2}}.
\end{align}
\label{eq:chiemxyRsi}
\end{subequations}
We can clearly see the there is no angular invariance in the $xz$ plane.
\subsection{Angle invariant circular polarization converter}
To derive the equations for circular polarization and extract \( \te{T}\@{CP} \) and \( \te{R}\@{CP} \), we start with the following expressions for the transmitted and reflected electric fields in the linear and circular polarization basis:
\begin{align}
    \begin{bmatrix} 
    E\@{TM}^\text{t} \\ 
    E\@{TE}^\text{t}
    \end{bmatrix} 
    &= \te{T}\@{LP}
    \begin{bmatrix} 
    E\@{TM}^\text{i} \\ 
    E\@{TE}^\text{i} 
    \end{bmatrix} \\
    \begin{bmatrix} 
    E\@{RCP}^\text{t} \\ 
    E\@{RCP}^\text{t}
    \end{bmatrix} 
    &=\te{T}\@{CP}
    \begin{bmatrix} 
    E\@{RCP}^\text{i} \\ 
    E\@{RCP}^\text{i} 
    \end{bmatrix}
\end{align}
We can also write the circular polarization components as a function of the linear polarization components such as,
\begin{align}
    \begin{bmatrix} 
    E\@{RCP}^\text{i}\\ 
    E\@{RCP}^\text{i}
    \end{bmatrix} 
    &= \te{\Lambda}_1
    \begin{bmatrix} 
    E\@{TM}^\text{i} \\ 
    E\@{TE}^\text{i} 
    \end{bmatrix} \\
    \begin{bmatrix} 
    E\@{RCP}^\text{t}\\ 
    E\@{RCP}^\text{t}
    \end{bmatrix} 
    &= \te{\Lambda}_2
    \begin{bmatrix} 
    E\@{TM}^\text{t} \\ 
    E\@{TE}^\text{t} 
    \end{bmatrix}\\
    \begin{bmatrix} 
    E\@{RCP}^\text{r}\\ 
    E\@{RCP}^\text{r}
    \end{bmatrix} 
    &= \te{\Lambda}_3
    \begin{bmatrix} 
    E\@{TM}^\text{r} \\ 
    E\@{TE}^\text{r} 
    \end{bmatrix}
\end{align}
Same procedure could be applied to the reflection of the linear and polarization components. Then, by combining all the equations above, we reach the following transformation equation,
\begin{align}
    \ve{E}\@{CP}^\text{T} &= \underbrace{\te{\Lambda}\@{2} \cdot\; \te{T}\@{LP} \cdot\; \te{\Lambda}\@{1}^{-1}}_{\te{T}\@{CP}} \ve{E}\@{CP}^\text{I}, \\
    \ve{E}\@{CP}^\text{R} &= \underbrace{\te{\Lambda}\@{3} \cdot\; \te{R}\@{LP} \cdot\; \te{\Lambda}\@{1}^{-1}}_{\te{R}\@{CP}} \cdot \ve{E}\@{CP}^\text{I},
\end{align}
These equations can be expressed in $2 \times 2$ matrix form as follows:
\begin{equation}
\begin{split}
    \te{T}\@{CP} =& 
    \frac{1}{2}
    \begin{bmatrix}
        \left( T_{\text{TM/TM}} + T_{\text{TE/TE}} \right) &  
        \left( T_{\text{TM/TM}} - T_{\text{TE/TE}} \right) \\
        \left( T_{\text{TM/TM}} - T_{\text{TE/TE}} \right) & 
        \left( T_{\text{TM/TM}} + T_{\text{TE/TE}} \right)
    \end{bmatrix}\\
        &+\frac{j}{2}
    \begin{bmatrix}
        \left( T_{\text{TM/TE}} - T_{\text{TE/TM}} \right)  &  
        - \left( T_{\text{TM/TE}} + T_{\text{TE/TM}} \right) \\
        \left( T_{\text{TM/TE}} + T_{\text{TE/TM}} \right) & 
        -\left( T_{\text{TM/TE}} - T_{\text{TE/TM}} \right)
    \end{bmatrix}
\end{split}
\end{equation}
\begin{equation}
\begin{split}
    \te{R}\@{CP} =& 
    \frac{1}{2}
    \begin{bmatrix}
        \left( R_{\text{TM/TM}} - R_{\text{TE/TE}} \right) &  
        \left( R_{\text{TM/TM}} + R_{\text{TE/TE}} \right) \\
        \left( R_{\text{TM/TM}} + R_{\text{TE/TE}} \right) & 
        \left( R_{\text{TM/TM}} - R_{\text{TE/TE}} \right)
    \end{bmatrix}\\
    &+\frac{j}{2}
    \begin{bmatrix}
       \left( R_{\text{TM/TE}} + R_{\text{TE/TM}} \right)&  
        - \left( R_{\text{TM/TE}} - R_{\text{TE/TM}} \right) \\
        \left( R_{\text{TM/TE}} - R_{\text{TE/TM}} \right) & 
        - \left( R_{\text{TM/TE}} + R_{\text{TE/TM}} \right)
    \end{bmatrix}    
\end{split}    
\end{equation}
If we desire cross-polarization in the circular polarization case, then the diagonal elements of both \( T_{\text{CP}} \) and \( R_{\text{CP}} \) must be zero. This implies that \( T_{\text{TM/TM}} = -T_{\text{TE/TE}} \) and \( R_{\text{TM/TM}} = R_{\text{TE/TE}} \). For a structure characterized solely by the electromagnetic susceptibilities \( \chi_\text{em}^{yx} \) and \( \chi_\text{em}^{xy} \), the co-polarized equations are given by:
\begin{align}
    T_{\text{TM/TM}}(\theta) &= \frac{4 + k^2 \chi_\text{em}^{yx2}}{4 - k^2 \chi_\text{em}^{yx2}} \\
    T_{\text{TE/TE}}(\theta) &= \frac{4 + k^2 \chi_\text{em}^{xy2}}{4 - k^2 \chi_\text{em}^{xy2}} \\
    R_{\text{TM/TM}}(\theta) &= \frac{4 j k \chi_\text{em}^{xy}}{-4 + k^2 \chi_\text{em}^{xy2}} \\
    R_{\text{TE/TE}}(\theta) &= \frac{4 j k \chi_\text{em}^{yx}}{4 - k^2 \chi_\text{em}^{yx2}}
\end{align}
In the case where no cross-polarization is present in the linear domain, the expressions simplify to:
\begin{align}
    \te{T}\@{CP} &= 
    \frac{1}{2}
    \begin{bmatrix}
        \left( T_{\text{TM/TM}} + T_{\text{TE/TE}} \right) &  
        \left( T_{\text{TM/TM}} - T_{\text{TE/TE}} \right) \\
        \left( T_{\text{TM/TM}} - T_{\text{TE/TE}} \right) & 
        \left( T_{\text{TM/TM}} + T_{\text{TE/TE}} \right)
    \end{bmatrix}
\end{align}
\begin{align}
    \te{R}\@{CP} &= 
    \frac{1}{2}
    \begin{bmatrix}
        \left( R_{\text{TM/TM}} - R_{\text{TE/TE}} \right) &  
        \left( R_{\text{TM/TM}} + R_{\text{TE/TE}} \right) \\
        \left( R_{\text{TM/TM}} + R_{\text{TE/TE}} \right) & 
        \left( R_{\text{TM/TM}} - R_{\text{TE/TE}} \right)
    \end{bmatrix}
\end{align}
To satisfy the circular polarization condition, the following relationship must hold:
\begin{equation}
    \chi_\text{em}^{yx} \chi_\text{em}^{xy} = \pm \frac{4}{k^2}
\end{equation}
\subsection{Binary-Phase Angle-Invariant Pseudochirality}
In this case we consider the following susceptibilities
\begin{align}
\begin{aligned}
    \te{\chi}_\text{ee} &= 0, \quad &\te{\chi}_\text{mm} &= 0, \\
    \te{\chi}_\text{em} &=
    \begin{pmatrix}
        0 & 0 & 0 \\
        0 & 0 & \chid{em}{yz} \\
        0 & \chid{em}{zy} & 0
    \end{pmatrix}, \quad &
    \te{\chi}_\text{me} &= 
    \begin{pmatrix}
        0 & 0 & 0 \\
        0 & 0 & \chid{me}{yz} \\
        0 & \chid{me}{zy} & 0
    \end{pmatrix} .
\end{aligned}
\end{align}

Then the given TM transmission and reflection coefficients are reduced to,
\begin{align}
R_{\text{co}}(\theta, \phi) &= \frac{4 k_y^2 \left(\chid{em}{yz} - \chid{em}{zy}\right)\left(\chid{em}{yz} + \chid{em}{zy}\right)}{
\left(-4 + k_y^2 \chid{em}{yz^2}\right) \left(-4 + k_y^2 \chid{em}{zy^2}\right)} \\
T_{\text{co}}(\theta, \phi) &= \frac{\left ( 4 - k_y^2 \chid{em}{yz} \chid{em}{zy} \right ) \left ( 4 + k_y^2 \chid{em}{yz} \chid{em}{zy} \right )}{
\left(-4 + k_y^2 \chid{em}{yz^2}\right) \left(-4 + k_y^2 \chid{em}{zy^2}\right)} \\
R_{\text{cr}}(\theta, \phi) &= \frac{2j k_y \left(\chid{em}{yz} - \chid{em}{zy}\right) 
\left(4 + k_y^2 \chid{em}{yz} \chid{em}{zy}\right)}{
\left(-4 + k_y^2 \chid{em}{yz^2}\right) \left(-4 + k_y^2 \chid{em}{zy^2}\right)} \\
T_{\text{cr}}(\theta, \phi) &= -\frac{2j k_y \left(\chid{em}{yz} + \chid{em}{zy}\right) 
\left(-4 + k_y^2\chid{em}{yz} \chid{em}{zy}\right)}{
\left(-4 + k_y^2 \chid{em}{yz^2}\right) \left(-4 + k_y^2 \chid{em}{zy^2}\right)}
\end{align}
And if the susceptibilities are equal to each other, meaning that,
\begin{align}
    \label{xemyzxemze}
    \chid{em}{zy} = \chid{em}{yz},
\end{align}
Then, the reflection coefficients are null, and the transmission coefficients are reduced to,
\begin{subequations}
\label{eq:chiemyzTR}
\begin{align}
    T\@{co}(\theta, \phi) &= \frac{4 + k_y^2 \chi\@{em}^{{yz}^2}}{4 - k_y^2 \chi\@{em}^{{yz}^2}}, \label{eq:chiemyzT} \\
    T\@{cr}(\theta, \phi) &= \frac{4jk_y \chi\@{em}^{yz}}{4 - k_y^2 \chi\@{em}^{{yz}^2}}.
\end{align}
\label{eq:chiemyzR}
\end{subequations}

\newpage
\renewcommand{\theequation}{D\arabic{equation}}
\section{\label{sec_structure_dim}Simulation for dog-bone structure}
\begin{figure}[H]
    \centering
    \includegraphics[width=0.5\linewidth]{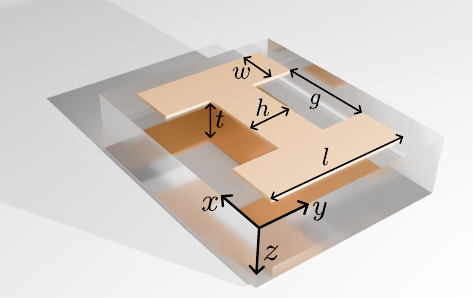}
    \caption{\textbf{Anisotropic unit cell, double dog bone structure for microwave simulation.}}
    \label{bone_structure1}
\end{figure}

The dog-bone structure as illustrated in \figref{bone_structure1}{} is a three-dimensional unit cell that represents certain symmetries. The symmetries 

It can be easily seen that the present symmetries are the mirror symmetries along the Cartesian axis and also the $180^\circ$ rotation along the same axis. Those symmetries induce the off-diagonal elements of the $\te{\chi}_{\text{ee}}$ and $\te{\chi}_{\text{mm}}$ tensors to be null. On the other hand, the magnetic-to-electric and electric-to-magnetic tensors, $\te{\chi}_{\text{em}}$ and $\te{\chi}_{\text{me}}$, respectively, are null tensors. If we excite the element with a TM-polarized field, the tensor elements $\chi_{\text{ee}}^{xx}$, $\chi_{\text{ee}}^{zz}$, and $\chi_{\text{mm}}^{yy}$ are only excited.
The dimensions of this structure are \mbox{$l = 3$~mm}, $g = 2.5$~mm, $h = 0.5$~mm, $w = 1$~mm, $t = 1$~mm. The simulation in the main text is obtained on CST Studio Suite 2023.
\begin{figure}[H]
    \centering
    \includegraphics[width=0.5\linewidth]{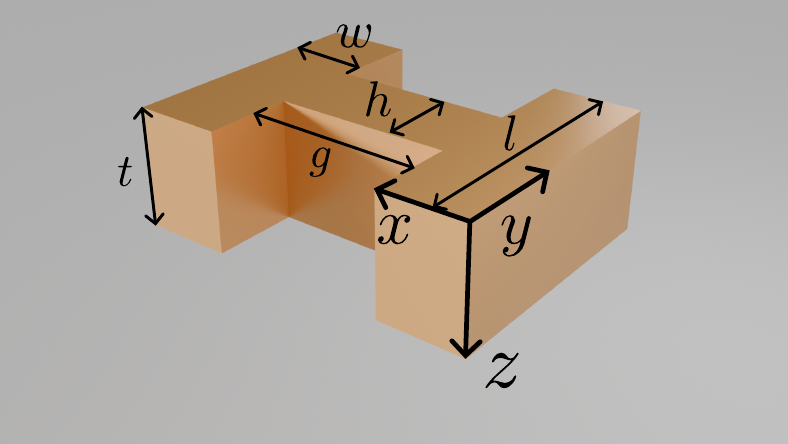}
    \caption{\textbf{Anisotropic unit cell, H shaped structure for optical frequency simulation.}}
    \label{bone_structure2}
\end{figure}

For the optical domain dog bone structure, the geometrical parameters are : 
$t = 400$~nm, $l = 400$~nm, $g = 160$~nm, $w = 90$~nm and $h= 175$~nm.
\begin{figure}[H]
    \centering
    \includegraphics[width=0.5\linewidth]{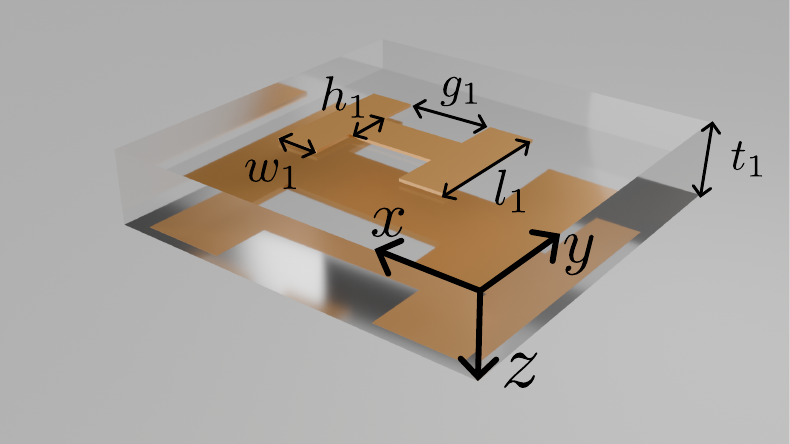}
    \caption{\textbf{Bianisotropic unit cell, double dog bone structure for bianisotropic susceptibilities in microwave simulation.}}
    \label{bone_structure3}
\end{figure}
For the microwave domain, the bianisotropic dog bone structure have different structure for the top and bottom structures, and the geometrical parameters are : 
$t_1 = 1.52$~mm, $l_1 = 2.5$~mm, $g_1 = 4$~mm, $w_1 = 0.5$~mm and $h_1= 0.5$~mm, $l_2 = 4.5$~mm, $g_2 = 4$~mm, $w_2 = 0.5$~mm and $h_2= 0.5$~mm.

\newpage
\renewcommand{\theequation}{F\arabic{equation}}
\section{\label{sec_homo}Homogenization of the susceptibilities}
\subsection{Diagonal anisotropic metasurface method}
This first method suggests to use \eqqref{eq_1} to find the susceptibilities. The number of susceptibilities is three, i.e. $\chid{ee}{xx}$, $\chid{ee}{zz}$ and $\chid{mm}{yy}$. The first concern here is that we only have two equations and two unknowns. If we look carefully to the equation, we can notice that for $k_x = 0$ (thus $k_z = k_0$), the susceptibility $\chid{ee}{zz}$ disappears from the transmission and reflection coefficients, letting only two unknowns. Thus, by inverting the equations, we can obtain,
\begin{align}
    \chid{ee}{xx} \Big|_{k_x = 0} &= \frac{2j(R + T - 1)}{k_z(R + T + 1)} \\
    \chid{mm}{yy} \Big|_{k_x = 0} &= \frac{2jk_z(R - T + 1)}{k^2(R - T + 1)}
\end{align}
Once obtain the susceptibilities $\chid{ee}{xx}$ and $\chid{mm}{yy}$, we can proceed and find $\chid{ee}{zz}$. In fact, know we have two equations and one unknown, letting the system being overdetermined. In order to solve this system we can use the least square method.
Let the following system being an overdetermined, noted as,
\begin{align}
    \mathbf{Ax} = \mathbf{b}.
\end{align}
Where $\mathbf{A}$ is a $m \times n$ matrix ($m > n$), $\mathbf{x}$ is an $n\times1$ vector unknowns and $\mathbf{b}$ is an $m\times1$ vector of observed values. The goal is to find $\mathbf{x}$ that minimizes the error, defined as the Euclidean norm of the residual vector $r = \mathbf{Ax - b}$. This can be formulated as minimizing the sum of the squared residuals:
\begin{align}
    \min_{\mathbf{x}} \| \mathbf{Ax - b} \|
\end{align}
The solution to this minimization problem can be found by setting the derivative of the sum of the squared residuals to zero, leading to the normal equations,
\begin{align}
    \mathbf{x} = \mathbf{\left( A^T A\right)^{-1} A^T b}
\end{align}
The algorithm is applied for a set of data {$t_{\theta_i},r_{\theta_i}$} for a given incident angle $\theta_i$. 
\begin{align}
    \mathbf{A_{\theta_i}x_{\theta_i}} = \mathbf{b_{\theta_i}}.
\end{align}
Each iteration, the obtained vector $x_{\theta_i}$ should be used to calculate the cost function $F_{MSE}$ which is defined as the Mean Square Error (MSE), given by
\begin{align}
    F_\text{MSE} = \frac{1}{2N} \sum_k \left [ \left( T_{\theta_k} - T_{\theta_j}\right)^2 + \left( R_{\theta_k} - R_{\theta_j}\right)^2 \right ].
\end{align}
Then the solution $x_{\theta_j}$ that give the lowest error will be chosen as the homogenization angle.
\subsection{Bianisotropic metasurface}
In this case, the homogenization is more accurate since we have three unknowns ($\chid{ee}{xx}$, $\chid{mm}{yy}$ and $\chid{em}{zz}$), but also three equations: the forward transmission and reflection ($T\@{co}^{+}, R\@{co}^{+}$), and the backward reflection ($R\@{co}^{-}$). In this case, $T\@{co}^{+} = T\@{co}^{-}$ and $R\@{co}^{+} \neq R\@{co}^{-}$, leading to a solvable system. By solving the co-transmission and co-reflection in forwards propagation and the co-reflection in backwards propagation, we obtain 
\begin{align}
    R\@{co}^{+}(\theta) &= \frac{-2k_z^2 \chi_\text{ee}^{xx} - 4k k_z \chi_\text{em}^{xy} + 2k^2 \chi_\text{mm}^{yy}}
    {2k_z^2 \chi_\text{ee}^{xx} + 2k^2 \chi_\text{mm}^{yy} + j k_z \left[-4 + k^2 \left(\chi_\text{em}^{xy^2} + \chi_\text{ee}^{xx} \chi_\text{mm}^{yy}\right)\right]}, \\
    T\@{co}^{+}(\theta) = T\@{co}^{-}(\theta) &= -\frac{j k_z \left[4 + k^2 \left(\chi_\text{em}^{xy^2} + \chi_\text{ee}^{xx} \chi_\text{mm}^{yy}\right)\right]}
    {2k_z^2 \chi_\text{ee}^{xx} + 2k^2 \chi_\text{mm}^{yy} + j k_z \left[-4 + k^2 \left(\chi_\text{em}^{xy^2} + \chi_\text{ee}^{xx} \chi_\text{mm}^{yy}\right)\right]}, \\
    R\@{co}^{-}(\theta) &= \frac{-2k_z^2 \chi_\text{ee}^{xx} + 4k k_z \chi_\text{em}^{xy} + 2k^2 \chi_\text{mm}^{yy}}
    {2k_z^2 \chi_\text{ee}^{xx} + 2k^2 \chi_\text{mm}^{yy} + j k_z \left[-4 + k^2 \left(\chi_\text{em}^{xy^2} + \chi_\text{ee}^{xx} \chi_\text{mm}^{yy}\right)\right]}.
\end{align}
As it can be observe, the forwards and backwards co-reflection coefficients are not the same, i.e. $R\@{co}^{+} \neq R\@{co}^{-}$, thus, we can solve the system
\begin{align}
    \chi_\text{ee}^{xx} &= \frac{2j \left[1 - \left(R\@{co}^{-} + R\@{co}^{+}\right) + \left(R\@{co}^{-} R\@{co}^{+}\right) - T\@{co}^{+2} \right]}
    {k_z \left[\left(R\@{co}^{-} R\@{co}^{+}\right) - \left(1+2T\@{co}^{+} + T\@{co}^{+2}\right)\right]}, \\
    \chi_\text{mm}^{yy} &= \frac{2jk_z \left[1 + \left(R\@{co}^{-} + R\@{co}^{+}\right) + \left(R\@{co}^{-} R\@{co}^{+}\right) - T\@{co}^{+2} \right]}
    {k^2 \left[\left(R\@{co}^{-} R\@{co}^{+}\right) - \left(1+2T\@{co}^{+} + T\@{co}^{+2}\right)\right]}, \\
    \chi_\text{em}^{xy} &= \frac{2j \left[R\@{co}^{-} - R\@{co}^{+}\right]}
    {k \left[\left(R\@{co}^{-} R\@{co}^{+}\right) - \left(1+2T\@{co}^{+} + T\@{co}^{+2}\right)\right]}.
\end{align}
From here, we can extract the transmission and reflection from the simulation software to compute the susceptibilities.

\end{document}